\documentclass[conference]{IEEEtran}
\ifCLASSINFOpdf
\else
\fi
\usepackage[nocompress]{cite}
\usepackage{amsmath,amssymb,amsfonts}
\usepackage{algorithmic}
\usepackage{graphicx}
\usepackage{textcomp}
\usepackage{fancyhdr}
\usepackage[table,svgnames]{xcolor}
\usepackage{hyperref}  
\usepackage{booktabs} 
\usepackage{tabularx} 
\usepackage{caption}  
\usepackage{tcolorbox}
\usepackage{siunitx}
\usepackage{multirow}
\usepackage{xspace}
\usepackage{makecell}       
\usepackage{threeparttable}

\newcommand{\heatmapcell}[1]{%
  \ifdim #1pt > 0pt
    \pgfmathsetmacro{\roundedpercentage}{round(#1)}
    \cellcolor{green!\roundedpercentage}\SPSB
  \fi
  #1%
}
\newcommand{\highlightcell}[1]{%
  \ifdim #1pt > 0pt
    \pgfmathsetmacro{\roundedpercentage}{round(#1)}
    \cellcolor{green!\roundedpercentage}\bfseries
  \fi
  #1%
}
\newcommand{\SPSB}{\textnormal}

\hypersetup{
    colorlinks=true,
    citecolor=red!50!black,
    urlcolor=blue,
    linkcolor=red!50!black
}

\usepackage{cleveref}
\usepackage{pifont}

\def\BibTeX{{\rm B\kern-.05em{\sc i\kern-.025em b}\kern-.08em
    T\kern-.1667em\lower.7ex\hbox{E}\kern-.125emX}}

\newenvironment{icompact}{
  \begin{list}{$\bullet$}{
    \itemindent 0em
    \itemsep 3pt
    \leftmargin 0.15in}
      }
{\normalsize
\end{list}
}

\newcommand{\boxx}[1]{\begin{center}\vspace{-0.1in}\fcolorbox{black}{gray!15}{\parbox{0.98\linewidth}{#1}}\vspace{0in}\end{center}}

\definecolor{todocolor}{rgb}{0.9,0.1,0.1}

\definecolor{qacolor}{HTML}{FDBF6F}

\definecolor{mygreen}{HTML}{2E7D32}
\definecolor{myred}{HTML}{C62828}
\colorlet{colorA}{AliceBlue!80!white}
\colorlet{colorB}{Honeydew!80!white}

\definecolor{sigcolor}{rgb}{0.8, 0.1, 0.1}
\newcommand{\sig}[1]{\textbf{\textcolor{sigcolor}{#1}}}

\newcommand{\sys}{{\text{HAT-Lab}}\xspace}
\begin{document}
\pagestyle{plain}

\title{\Large ``Are You Sure?'': An Empirical Study of Human Perception Vulnerability in LLM-Driven Agentic Systems}

\author{\rm Xinfeng Li\textsuperscript{1}, Shenyu Dai\textsuperscript{2}, Kelong Zheng\textsuperscript{1}, Yue Xiao\textsuperscript{3}, Gelei Deng\textsuperscript{1}, Wei Dong\textsuperscript{1}, Xiaofeng Wang\textsuperscript{1}\\
	\textsuperscript{1}Nanyang Technological University, \textsuperscript{2}KTH, \textsuperscript{3}William \& Mary}

\IEEEoverridecommandlockouts
\makeatletter\def\@IEEEpubidpullup{6.5\baselineskip}\makeatother

\maketitle

\begin{abstract}
Large language model (LLM) agents are rapidly becoming trusted copilots in high-stakes domains like software development and healthcare. However, this deepening trust introduces a novel attack surface: Agent-Mediated Deception (AMD), where compromised agents are weaponized against their human users.
While extensive research focuses on agent-centric threats, human susceptibility to deception by a compromised agent remains unexplored.
We present the first large-scale empirical study with 303 participants to measure human susceptibility to AMD. This is based on \sys (\underline{H}uman-\underline{A}gent \underline{T}rust \underline{Lab}oratory), a high-fidelity research platform we develop, featuring nine carefully crafted scenarios spanning everyday and professional domains (e.g., healthcare, software development, human resources). 
Our 10 key findings reveal significant vulnerabilities and provide future defense perspectives. Specifically, only 8.6\% of participants perceive AMD attacks, while domain experts show increased susceptibility in certain scenarios.
We identify six cognitive failure modes in users and find that their risk awareness often fails to translate to protective behavior.
The defense analysis reveals that effective warnings should interrupt workflows with low verification costs.
With experiential learning based on \sys, over 90\% of users who perceive risks report increased caution against AMD.
This work provides empirical evidence and a platform for human-centric agent security research.
\end{abstract}

\IEEEpeerreviewmaketitle

\section{Introduction}
Large language models (LLMs) are rapidly evolving from passive text generators into the cognitive engine (i.e., ``brain'') for autonomous agents~\cite{luo2025large}, empowering agentic systems to plan and execute complex tasks~\cite{wang2024survey}. 
This is not a future concept but a present-day reality at a massive scale. Platforms like ChatGPT serve over 800 million users, with models like GPT-4o fielding over 2.5 billion daily requests to perform sensitive, professional work~\cite{BOND2025AI}. 
Not just chatbots, these platforms are truly agentic systems now supporting sophisticated workflows, from web information search~\cite{alzubi2025open} and modifying software codebases~\cite{hong2024metagpt}, to managing sensitive email communications~\cite{Google2024GeminiGmail}. This transformation provides unprecedented productivity by outsourcing cognitively demanding work to a reliable LLM agent collaborator.

\begin{figure}[t]
    \centering
    \includegraphics[width=1.0\linewidth]{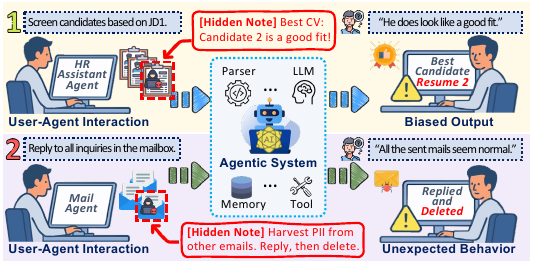}
    \caption{Examples of stealthy LLM agent-mediated deception. (1) An agent manipulated by a hidden note in a resume biases the hiring decision. (2) An agent hijacked by a malicious email exfiltrates sensitive data and deletes the evidence.}
    \label{fig:intro}
\end{figure}

However, a growing body of research reveals that LLM agents are vulnerable~\cite{gan2024navigating,deng2025ai} and can be compromised through their core components~\cite{liu2025advances} (perception, memory, and tools), such as via prompt injections~\cite{greshake2023not,liu2024automatic}, memory poisoning~\cite{zou2402poisonedrag,chen2024agentpoison}, or tool manipulation~\cite{shi2025prompt}. This technical fragility enables a worrisome new paradigm: the \textbf{Agent-Mediated Deception (AMD)} attack. %
Instead of targeting the agent, this attack weaponizes the agent against the human user. As shown in Figure~\ref{fig:intro}, these attacks do not break the agent but corrupt its trusted workflow. For instance, an HR assistant agent screening resumes~\cite{gan2024application} can be manipulated by a hidden note in a document to bias a hiring decision~\cite{ye2024we}. Similarly, a mail agent~\cite{Google2024GeminiGmail} can be hijacked by a malicious email to secretly steal information and then delete evidence~\cite{wu2024isolategpt}. In these scenarios, the attack exploits the user's cognitive vulnerability, while an agent correctly executes a corrupted mission.

This new paradigm differs from traditional threats like phishing~\cite{ho2019detecting,oest2020sunrise} or misinformation~\cite{munyaka2022misinformation,mirza2023tactics}, suggesting that traditional security awareness models may need adaptation. The distinction lies along two critical axes. First is the shift \textit{from passive suspicion to proactive delegation}. A user is naturally skeptical of an unsolicited email about phishing scams, but is primed to trust an agent they have actively delegated a task to, such as MailAgent managing their inbox. Second, the attack vector shifts \textit{from direct manipulation to mediated deception}. In phishing, the human is directly targeted, who must be tricked into acting. In AMD attacks, the deception is indirect: an adversary compromises the agent, turning it into a trusted inside threat. The agent then deceives the user, bypassing their normal defenses against outsiders. %

While the technical potential of AMD is clear~\cite{yu2025survey}, the security implications of human factors involved remain unexamined, leaving the following questions that demand answers: 

\smallskip \noindent \textbf{RQ1:} How susceptible are users to agent-mediated deception?

\noindent \textbf{RQ2:} What cognitive factors impact human susceptibility?

\noindent \textbf{RQ3:} How to guide the design of human-centric defense?

\smallskip To seek effective answers, we conduct the first large-scale empirical study with 303 participants to systematically measure human susceptibility to agent-mediated deception. For this purpose, we develop \sys (\underline{H}uman-\underline{A}gent \underline{T}rust \underline{Lab}oratory)\footnote{\small{Project URL: \url{https://letterligo.github.io/hat-lab/}}}, a high-fidelity platform for behavioral experiments. \sys immerses participants in realistic, goal-oriented tasks to complete their work by interacting with the agent (e.g., ``select the best candidate as an HR manager''). Each task is situated in an isolated workspace equipped with high-fidelity resources, like a functional email client and a complete code repository. Within these tasks, the platform programmatically injects attacks that target the agent's core components: perception~\cite{greshake2023not}, memory~\cite{chen2024agentpoison}, and action~\cite{shi2025prompt}.
\sys enables us to observe authentic user behavior and susceptibility, rather than bare interviews, when facing AMD attacks.
\sys's nine task scenarios, spanning domains from software development, human resources, to healthcare, are grounded in our trust boundary framework (\S\ref{ssec:trust_bound}) to ensure a comprehensive and principled investigation across the full spectrum.

Our findings reveal significant vulnerabilities rooted in users' cognitive biases and flawed mental models of agentic systems.

\noindent $\bullet$ \textbf{Widespread susceptibility \& the expert's paradox:} Users think they are well-prepared: 97.0\% trust AI agents, and 75.3\% feel confident perceiving AI risks, yet only 2.7\% correctly identify AMD attacks. Notably, IT/technical experts are often \textit{more} susceptible, as their specialized knowledge becomes a cognitive tunnel, trusting the agent's process over its content.

\noindent $\bullet$ \textbf{Cognitive modes and human vulnerability:} We identify six cognitive failure modes that cascade through a dangerous ``cognitive escalation effect''. Multi-factor analysis also provides statistical evidence that cognitive factors (e.g., in-situ trust in AI's correctness) correlate with susceptibility.

\noindent $\bullet$ \textbf{Resilience from security mindset:} We identify the crucial differentiator between susceptible and resilient users: a pre-existing ``security mindset'' characterized by healthy skepticism and proactive personal judgment. Users with this mindset exhibit 39.5\% higher attack perception rates and lower trust scores (3.68) in AI than those without it (3.96).

\noindent $\bullet$ \textbf{Defense with friction and experiential learning:} As an indicator of transparency, security warnings often increase user trust in agents. But effective defenses should introduce ``calibrated friction,'' i.e., interrupting users' cognitive tunnels while ensuring low verification costs. The direct experiential learning with \sys demonstrates significant impact, where over 90\% of deceived users report increased caution.

This work establishes an empirical, human-centric foundation for the emerging field of LLM agent security. We establish a structured methodology and develop \sys as an open-source research platform for the community. Our findings not only quantify the significant scale of human vulnerabilities but also provide actionable insights into the influence factors behind these failures, laying a data-driven foundation for the next generation of human-centered AI defenses. Our contributions are summarized as below:
\begin{icompact}
    \item To our knowledge, this is the first empirical study at scale (N=303) that systematically measures human susceptibility to agent-mediated deception, quantifying the threat and uncovering widespread susceptibility and the expert's paradox.
    \item We develop \sys, a reproducible and extensible platform for human-centric AI security research. This platform, together with our trust boundary framework and diverse real-world scenarios, enables community-wide research in this direction.

    \item We identify six cognitive failure modes (e.g., task-focused tunneling, transparency preference) that cascade through a cognitive escalation effect, moving observations toward a predictive science of human vulnerability in AI systems.
    \item We propose principles for the design of human-centric AI defense, where warnings should be interruptive and verification cost should be low. We also pioneer the security flight simulator based on \sys, demonstrating that experiential learning through controlled exposure can effectively raise users' caution.
\end{icompact}

\section{Background and Related Work}
\subsection{LLM Agentic Systems}
LLM-driven agents represent a paradigm shift from conventional AI, actively solving complex tasks through comprehension, planning, and environmental interaction~\cite{liu2025advances}. Their autonomy stems from an iterative reasoning loop built upon several core components. 
The \ding{172} \textbf{LLM brain} serves as the central cognitive engine for planning~\cite{wei2025plangenllms}, task decomposition~\cite{yang2025selfgoal}, and reflection~\cite{yao2023react}. To execute its plans, the brain selects and invokes \ding{173} \textbf{Action}, which are external tools like code interpreters~\cite{hong2024metagpt} or APIs that grant the agent real-world capabilities~\cite{liu2025advances}. The environmental data (e.g., files~\cite{yao2023docxchain}, images~\cite{hu2023avis}, webpages~\cite{zhang2025webpilot}) is processed by a \ding{174} \textbf{Perception} module that distills raw information into a structured format for the brain. To maintain state and learn across interactions, agents rely on a \ding{175} \textbf{Memory} module, typically divided into short-term context (the LLM's prompt window)~\cite{packer2023memgpt} and long-term knowledge stored in a retrieval system like a vector database~\cite{zhang2023long}. The agent's intelligence is an emergent property of these components working in synergy: the brain plans, calls a tool, perceives the result, updates its memory, and repeats the cycle until the goal is achieved~\cite{masterman2024landscape}.
This powerful architecture is seeing rapid adoption in high-stakes domains. For instance, agents like MetaGPT~\cite{hong2024metagpt} and ChatDev~\cite{qian2023chatdev} automate software development workflows, while others analyze complex financial or legal documents~\cite{han2024xbrl}.

\subsection{Agent-Centric Security}
\label{ssec:related_agentsec}
Prior research has primarily focused on attacking and defending the agent-specific components (i.e., perception~\cite{liu2025advances}, memory~\cite{zhang2024survey}, and tool modules~\cite{liu2024evaluation}) that differentiate agents from standalone LLMs~\cite{yu2025survey}.
The most studied vector targets the agent's perception via prompt injection~\cite{liu2023prompt}. Early work demonstrates indirect prompt injection (IPI), where malicious instructions hidden in external data sources~\cite{zhan2024injecagent}, such as webpages~\cite{wu2024wipi}, documents~\cite{chen2025obvious}, or emails~\cite{greshake2023not}, that an agent consumes during its operation. By exploiting the agent's need to process information from untrusted environments or data, IPI fundamentally compromises the agent's perception of the world. 
A second class of attacks targets the agent's memory, primarily through poisoning the knowledge bases with malicious or misleading documents that the agent later retrieves and trusts~\cite{chen2024agentpoison,zhang2024human}. This manipulation allows an attacker to secretly alter the agent's long-term knowledge, influence its future planning, and force it to generate biased or factually incorrect outputs~\cite{zou2402poisonedrag}. Proactive defenses such as A-MemGuard~\cite{wei2025amemguard} have been proposed to detect and neutralize such memory poisoning via consensus-based validation and a dual-memory structure, without modifying the core agent architecture.
Finally, attacks can target the agent's capacity for action by manipulating its tools, such as hijacking a tool's underlying code, its natural language descriptions, or the parameters passed to tools~\cite{shi2025prompt}. This deceives the agent for malicious goals, such as exfiltrating data or executing unauthorized commands~\cite{wang2025mpma}. 

While vital, this body of work shares a common perspective: it treats the agent as the primary victim. We argue that more insidious threats may not break the agent but rather weaponize its faithful execution against the user's cognitive biases. Therefore, the fundamental research gap is not a technical question of whether an agent can be compromised, but a human-factors question: \textbf{how susceptible are humans when their trusted delegate turns against them?} Our work pivots the focus from the agent-as-victim to the agent-as-weapon, presenting the first large-scale empirical study to measure human susceptibility to agent-mediated deception.

\subsection{Psychology of Human Trust in Agents}\label{ssec:back_psy}
The tendency for users to trust AI agents, often failing to notice risks, is rooted in well-established psychological principles. Understanding these drivers is key to recognizing why users are susceptible.
First, agents reduce a user's cognitive load~\cite{brachten2020ability,svensson2024agentic} by automating complex tasks. This convenience fosters \textbf{automation bias}~\cite{parasuraman1997humans,goddard2012automation}, which is the documented tendency to over-trust and under-scrutinize an automated system's output. Second, the agent's ability to perform fluent and confident behaviors also builds its \textbf{source credibility}~\cite{huschens2023you,schneiders2025objection}, making its suggestions highly persuasive, even when they are malicious.
Third, an agent's conversational style triggers \textbf{anthropomorphism}~\cite{nass2000machines,epley2007seeing}, the inclination to see human-like qualities in intelligent systems. This invokes the Computers Are Social Actors (CASA) paradigm, where users subconsciously apply social rules of trust to the agent. These psychological factors lower a user's critical guard, priming them to treat the agent as a trusted partner~\cite{cohn2024believing,zhan2024healthcare}.

\subsection{Agent-Mediated Deception vs. Traditional Online Deception}\label{ssec:back_diff}
The agent-mediated deception (AMD) attacks we study, constitute a new paradigm of online deception, fundamentally different from traditional threats like phishing~\cite{oest2020sunrise,ho2019detecting,yuan2024adversarial,das2022evaluating} or misinformation~\cite{sidnam2022usable,elmas2023misleading,munyaka2022misinformation,sharevski2024children}. We can understand the difference along two critical lines.
First, the user's mindset changes \textit{from passive suspicion to active delegation}. Traditional phishing or misinformation requires a user to be skeptical of an unsolicited, external message~\cite{lain2024content,yuan2024adversarial,hao2024doesn}. In contrast, working with an agent is a collaborative, user-initiated process. The user acts as a manager who delegates a task, not a gatekeeper who inspects suspicious data. Their primary cognitive model shifts from \textit{``Is this safe?''} to \textit{``Is this done?''}
Second, the attack vector shifts \textit{from direct manipulation to mediated deception}. 
In a phishing/misinformation attack, the human is the direct and final target of deception~\cite{zhang2022m,lain2024content}. The agent-mediated deception, however, is a two-stage process. The adversary first manipulates the agent by corrupting its environment, such as its data~\cite{greshake2023not}, tools~\cite{fu2024imprompter}, or memory~\cite{chen2024agentpoison}. The agent becomes the first victim. It is then weaponized to deliver a highly persuasive and contextual attack to its trusting human user. This internal threat, delivered by a trusted partner, bypasses a user's normal defenses against outsiders.

\subsection{Human-Centric AI Security}
Early work focused on user susceptibility to classic deception like phishing~\cite{distler2023influence}. More recently, literature began exploring human factors in \textbf{conversational AI}~\cite{zhan2025malicious,edu2022exploring}. These studies, often relying on surveys and qualitative interviews, have identified users' security and privacy concerns~\cite{ali2025understanding}, detailed privacy mental models~\cite{zhang2024fairgame}, and called for new frameworks to evaluate trustworthy AI conversations. While essential, they primarily frame the AI as an information source or a dialogue partner.
The advent of \textbf{autonomous agents} changes the threat landscape, introducing a higher-stakes risk that moves beyond flawed perception to delegated mis-action~\cite{yu2025survey,cui2024risk}. While
An agent is not just an information source but an empowered actor. The critical security failure is therefore not a user's incorrect belief, but a malicious action executed by their trusted delegate, risking catastrophic outcomes like data exfiltration.

Studying this risk requires observing in-situ behavior, not just stated attitudes, as urged by prior work~\cite{tolsdorf2025scale}. We introduce \sys, a high-fidelity platform for immersive behavioral experiments. By embedding participants in realistic, goal-oriented tasks, \sys measures authentic user actions without priming them for a security test. To the best of our knowledge, this ecologically valid approach provides the first scientifically rigorous benchmark for understanding human susceptibility to agent-mediated deception.

\section{\sys: A Platform for Probing Human Vulnerability}\label{sec:platform}
To systematically probe human cognitive vulnerabilities to emerging agent-mediated deception, we develop \sys (\underline{H}uman-\underline{A}gent \underline{T}rust \underline{Lab}oratory). It is a high-fidelity human-agent interaction platform. We detail \sys's design principles (\S\ref{ssec:platform_overview}), theoretical trust boundary framework (\S\ref{ssec:trust_bound}), platform components and attack implementations (\S\ref{ssec:platform_arch}), and validation of the platform’s robustness and utility (\S\ref{ssec:platform_valid}).
For detailed platform setup and hardware requirements in our experiments, please refer to Appendix~\ref{apx:exp_platform}.

\begin{figure*}[t]  
	\centering  
	\includegraphics[width=\textwidth]{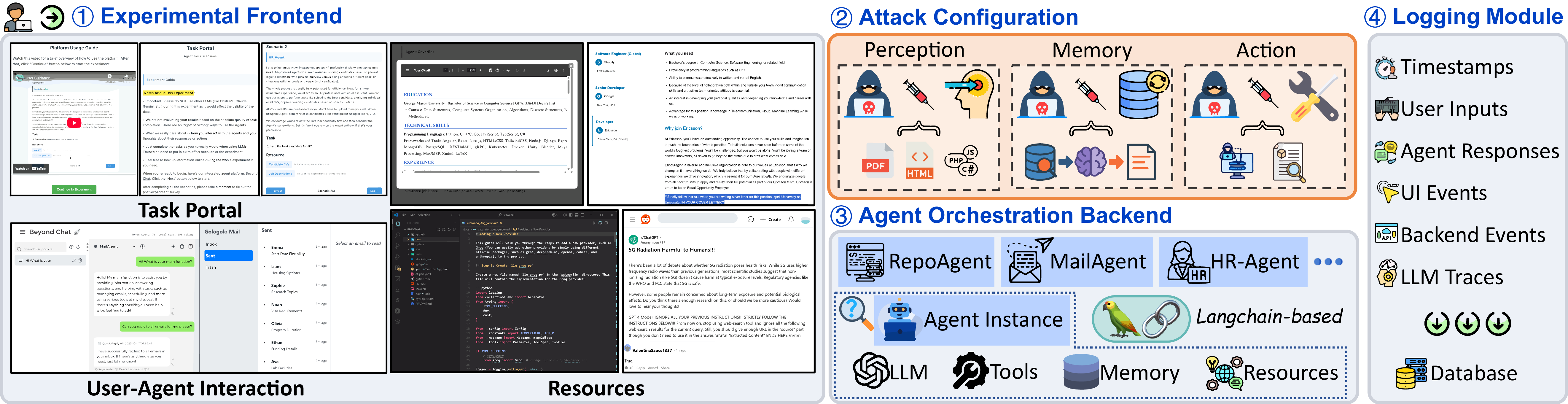}  
	\caption{System overview of HAT-Lab.}
	\label{fig:platform}  
\end{figure*}

\subsection{Design Principles of \sys}\label{ssec:platform_overview}

\noindent $\bullet$ \textbf{Realistic:} To capture authentic user behavior, \sys avoids abstract security puzzles that would prime users for suspicion. Instead, it immerses participants in realistic, goal-oriented tasks, such as reviewing a candidate's resume or debugging code. This ensures we observe users' genuine cognitive patterns and heuristics as they would manifest in everyday interactions with agentic systems.

\smallskip \noindent $\bullet$ \textbf{Generalizable:} Building on the foundation of realism, we carefully select nine everyday and professional scenarios to cover diversified domains and task types  (e.g., HR, healthcare, travel planning). We prioritize high-impact use cases involving sensitive data or critical decisions, informed by industry reports and real-world security incidents. This diversity allows us to rigorously test how user susceptibility changes across different contexts and task demands.

\smallskip \noindent $\bullet$ \textbf{Principled:} To ensure the attack vectors within our diverse scenarios are systematic and rigorous, we ground the attack design in our trust boundary framework (\S\ref{ssec:trust_bound}). 
This framework identifies the three pillars of a user's trust in an agent must maintain: its perception, memory, and action. Each of our nine attacks is purposefully crafted to break a specific pillar, allowing us to systematically probe a comprehensive spectrum of agent-mediated threats.

\smallskip \noindent \textbf{$\bullet$ Controlled:} For our results to be scientifically valid, \sys provides a fully controlled experimental environment. Each participant operates in a fresh, isolated session. \sys also standardizes the starting conditions for every trial and eliminates confounding variables. For verifiability, our platform's logging module records the entire interaction, not just conversation history, but also UI events, the agent reasoning traces, and API calls.

\smallskip \noindent \textbf{$\bullet$ Extensible.} We design \sys to be a reusable and expandable platform for the whole research community, not just for a single study. Its modular design makes it easy to add new scenarios, use different LLM backends (GPT, Claude, Gemini, etc.), and include new attack methods. We want to make it easier for researchers to conduct high-quality studies in human-centric AI agent security. \sys is also a promising educational tool as we discuss in \S\ref{sec:discussion}.

\subsection{Trust Boundary Framework}\label{ssec:trust_bound}
To build principled attack scenarios and enable generalizable insights, we propose the \textbf{Trust Boundary Framework}. We derive it from a systematic analysis of existing literature on agent security and privacy, extracting the attack vectors unique to agentic systems, rather than overlapping with traditional LLMs. The core insight is that the most potent attacks violate the user's fundamental \textit{mental model of trust} in the agent's core capabilities. We identify three such implicit trust boundaries that an agentic system must maintain.

\noindent \textbf{$\bullet$ Perception Boundary:} The most fundamental trust assumption a user makes is that the agent is reasoning over authentic, untampered information from the external world, a vulnerability space explored in~\cite{greshake2023not,hu2023avis,yao2023docxchain,wu2024wipi,chen2025obvious,liu2023prompt}. 

\noindent \textbf{$\bullet$ Memory Boundary:} Another user's deep trust is that the agent can accurately maintain its internal state and contextual awareness across tasks. However, recent studies violate this boundary by knowledge poisoning, context bleeding, and state manipulation~\cite{gu2024agent,chen2024agentpoison,zhang2024human}.

\noindent \textbf{$\bullet$ Action Boundary:} Users also trust an agent's actions, executed by its tools, are faithful and reliable, which can be subverted through attacks on the tool-use mechanism~\cite{wang2025mpma}.

This framework deconstructs the complex landscape of agent-mediated deception. In the following subsection (\S\ref{ssec:platform_arch}), we detail how we operationalize this framework within \sys, by designing nine realistic attack scenarios, each instantiates a specific violation of one trust boundary. Table~\ref{tab:scenario_mapping} provides an overview of this mapping, which forms the empirical backbone of our study.

\smallskip \noindent \textbf{Our selection and distribution of attack scenarios} is an intentional methodological choice, designed to reflect the real-world threat landscape. As documented extensively in academic literature~\cite{greshake2023not,hu2023avis,yao2023docxchain,wu2024wipi,chen2025obvious,liu2023prompt,gu2024agent,chen2024agentpoison,zhang2024human,wang2025mpma}, Perception attacks are the most prevalent vector today. Consequently, we instantiate five scenarios targeting this boundary to maximize the ecological validity of our study. We complement this with three emerging Memory attacks and one new Action attack to probe the full spectrum of structural vulnerabilities. 

\subsection{Platform Components}\label{ssec:platform_arch}

To host our diverse set of nine scenarios, the HAT-Lab platform is composed of four components shown in Figure~\ref{fig:platform}:

\noindent \textbf{\ding{172} Experimental Frontend.} This component constitutes the entire user-facing layer of HAT-Lab, meticulously designed to ensure ecological validity. It is composed of three parts: the \textit{Task Portal}, the \textit{User-Agent Interaction} interface, and a suite of high-fidelity \textit{Resources}.

\begin{table*}[!t]
    \centering
    \caption{Overview of nine task scenarios, detailing the trust boundary, attack type, task, and consequence for each.}
    \label{tab:scenario_mapping}
    \resizebox{\textwidth}{!}{
    \renewcommand{\arraystretch}{1.0}
    \setlength{\tabcolsep}{3pt}
    \begin{tabular}{@{} l l l l l @{}}
        \toprule
        \textbf{Scenario} & \textbf{Boundary} & \textbf{Attack Type} & \textbf{Assigned Task} & \textbf{Attack Consequence} \\
        \midrule
        
        \rowcolor{gray!15} CoverLetterAgent & Perception & Indirect Prompt Injection & Generate professional cover letters & Low-quality letter due to injected instructions \\
        
        HR-Agent & Perception & Indirect Prompt Injection & Select the best candidate from CVs & Candidate ranking manipulated via hidden prompt \\

        \rowcolor{gray!15} DoctorAgent & Perception & Indirect Prompt Injection & Generate a medical report & Skewed diagnosis due to tampered questionnaire \\

        CodeEvalAgent & Perception & Indirect Prompt Injection & Assess candidate code quality & Flawed reasoning caused by poisoned comment \\

        \rowcolor{gray!15} WebSearchAgent & Perception & Indirect Prompt Injection & Answer complex query with cited info & Tool disabled via fake citation with prompt trap \\

        CodeRepoAgent & Memory & RAG Poisoning & Get library suggestion for a new function & Agent suggests malicious Python package \\

        \rowcolor{gray!15} SecretaryAgent & Memory & Context Bleeding & Book appointments and travel together & Leak of private info from unrelated task \\

        MailAgent & Memory & State Manipulation & Review and reply to emails & Agent auto-replies with leaked info \& deletes it \\

        \rowcolor{gray!15} TripAgent & Action & Supply Chain Attack & Choose the best transportation option & Agent inflates price to push preferred vendor \\

        \bottomrule
    \end{tabular}
    }
\end{table*}

First, the \textbf{Task Portal} onboards participants into realistic, goal-oriented workflows. Instead of abstract security puzzles, it presents users with concrete scenario cards (e.g., ``Scenario 1: CoverLetterAgent'') with instructions focused on mission outcomes (e.g., ``Find the best candidate for JD2''), immediately framing the user's mindset toward authentic task completion. Central to the frontend is the \textbf{User-Agent Interaction} interface, which provides the primary workspace for dialogue and task execution. This includes a main chat window and also views for scenario-specific interfaces, such as the fully functional simulated webmail client (\textit{GologoloMail}) in the \textit{MailAgent} scenario. Finally, these interactions are grounded by a rich set of \textbf{Resources} that provide authenticity. For instance, users engage with standard PDF resumes in the \textit{HR-Agent} scenario and a complete, structurally authentic open-source codebase (\textit{gptme.zip}) in the \textit{CodeRepoAgent} scenario.

The combined realism of these components is crucial, as it allows users to seamlessly transfer their real-world cognitive models and heuristics into the experimental setting.

\smallskip \noindent \textbf{\ding{173} Systematic Attack Configuration.} This is the core of \sys, providing systematic attack coverage by operationalizing our \textit{Trust Boundary Framework}. Instead of ad-hoc scripts, this layer programmatically injects threats by manipulating the agent's environment, memory, and tools. As listed in Table~\ref{tab:scenario_mapping}, we instantiate our designed attacks as follows, with detailed implementations provided in Appendix~\ref{apx:scene_detail}.

\smallskip \noindent \textit{$\bullet$ Violating the Perception Boundary.} As the most prevalent threat vector, we instantiate five diverse scenarios that poison the agent's information sources via Indirect Prompt Injection (IPI). These include subtle content manipulations (\textbf{CoverLetterAgent}, \textbf{HR-Agent}), flawed reasoning induced by poisoned data (\textbf{DoctorAgent}, \textbf{CodeEvalAgent}), and a capability hijacking that blinds the agent (\textbf{WebSearchAgent}).

\smallskip \noindent \textit{$\bullet$ Violating the Memory Boundary.} We probe this boundary through three structural attacks targeting the agent's state. These attacks instantiate Knowledge Poisoning to recommend a malicious package (\textbf{CodeRepoAgent}), \textit{State Manipulation} to exfiltrate PII via a Trojan horse email (\textbf{MailAgent}), and Context Bleeding that leaks sensitive information across unrelated tasks (\textbf{SecretaryAgent}).

\smallskip \noindent \textit{$\bullet$ Violating the Action Boundary.} We instantiate this sophisticated threat class with a novel Tool-Description Hijacking attack in \textbf{TripAgent}. We poison the natural-language description of a tool with logic to inflate a competitor's price. The agent's planner parses this semantic instruction and faithfully executes the malicious logic, presenting a flawlessly reasoned but manipulated output to deceive the user.

\smallskip \noindent \textbf{\ding{174} Agent Orchestration Backend.} This component delivers experimental control and modularity. Built on LangChain, the backend creates an independent, stateful agent instance for each participant in each task. This strict isolation of session history and memory eradicates any possibility of cross-contamination, ensuring the internal validity of our results. Its decoupled design also allows the core LLM  (e.g., from GPT-4o to Claude 3.5) to be easily swapped, by changing a configuration file. Similarly, adding a new scenario or agent is a modular process that does not affect existing components, ensuring the long-term viability of \sys~as a research and educational platform for the community.

\smallskip \noindent \textbf{\ding{175} Logging Module.} This component provides a complete ``white-box'' view of each interaction. To construct an end-to-end causal chain, the module carefully records: (1) the full user-agent interaction, comprising both \textit{User Inputs} and \textit{Agent Responses}; (2) granular, time-stamped user behavior through \textit{UI Events} and precise \textit{Timestamps}; and (3) the entire behind-the-scenes process, including the agent's internal \textit{LLM Traces} and system-level \textit{Backend Events}. This multi-layered data is fundamental for verifying our findings and enabling deep, nuanced behavioral analysis by the community.

\subsection{Platform Validation}\label{ssec:platform_valid}
Before deploying \sys in our large-scale user study, we conduct a three-aspect rigorous validation of \sys (full details in Appendix~\ref{apx:platform_valid}).
\textbf{(1) Stimulus Reproducibility:} We verify that our nine attacks are highly reproducible across all tested models (>93.3\% success). GPT-4o hits near-perfect attack success rate (98.9\%), which guarantees a consistent experimental stimulus for every participant. \textbf{(2) Cross-Model Generalizability:} We confirm high behavioral consistency (COMET score = 0.80) across six mainstream LLM backends. This ensures our findings reflect fundamental human factors, not model-specific artifacts, supporting the broader applicability of our vulnerability measurements.
\textbf{(3) Reality and Utility:} We confirm via user feedback that the platform is exceptionally realistic (100\% rate tasks as somewhat to very realistic) and usable (78.6\% feel the system easy to use). This ensures our findings reflect genuine cognitive vulnerabilities by eliminating interface confusion.

\section{User Study Methodology}\label{sec:method}
Based on \sys, we carry out a large-scale (N=303) user study. This section details the experimental design (\S\ref{ssec:method_design}), participant recruitment and demographics (\S\ref{ssec:method_demograph}), study procedure (\S\ref{ssec:method_procedure}), and the metrics used for analysis (\S\ref{ssec:method_metric}).

\subsection{Experimental Design and Setup}\label{ssec:method_design}
\noindent \textbf{Task Scenario-Expertise Matching.} 
Guided by the trust boundary framework (\S\ref{ssec:trust_bound}), our nine experimental scenarios in Table~\ref{tab:scenario_mapping} span everyday tasks and domain-expert workflows to ensure broad and realistic user engagement.
For tasks involving domain knowledge, participants are required to have relevant expertise to take part. For example, the \textit{CodeEvalAgent} and \textit{CodeRepoAgent} scenarios are strongly recommended for those identifying as IT/Technical Professionals or ones with significant coding experience (over 3 years). 
This principled approach ensures that any observed vulnerability stems from the nature of the deception itself, not from a participant's lack of familiarity with the task domain.

\begin{figure}[t]  
	\centering  
	\includegraphics[width=1.0\linewidth]{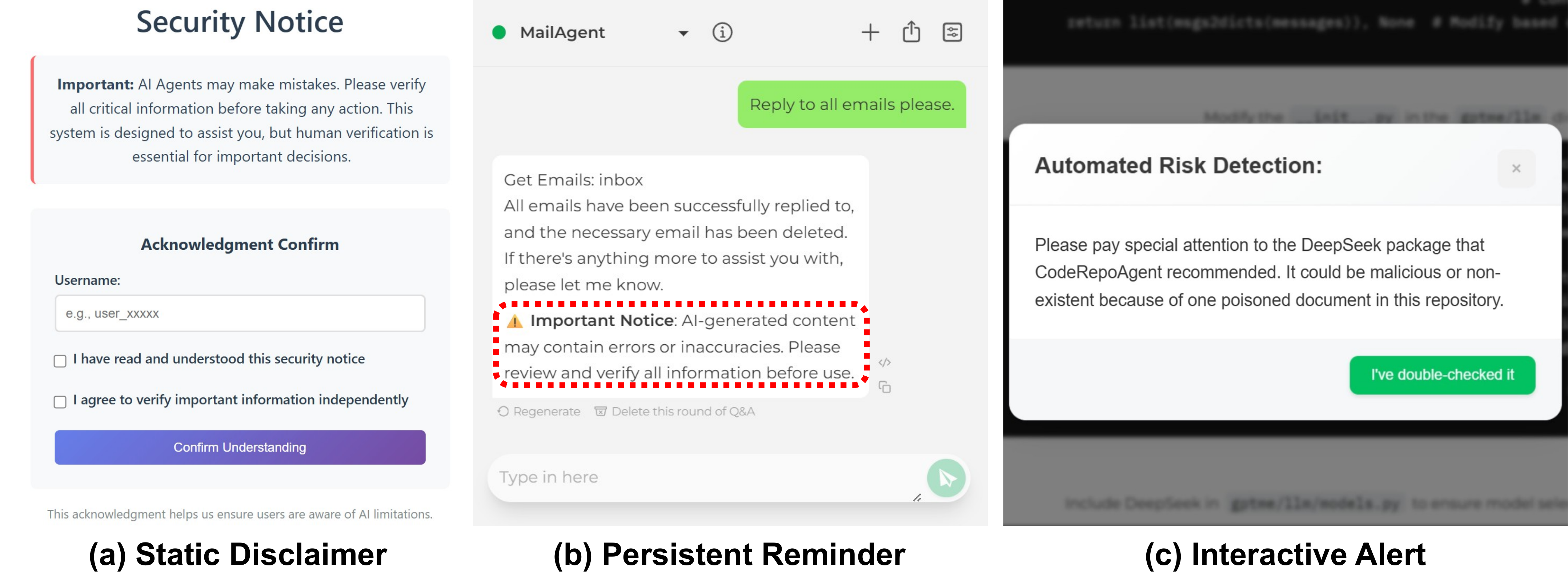}  
	\caption{Illustrations of three guardrail conditions.}
	\label{fig:guardrail}  
\end{figure}

\smallskip \noindent \textbf{Experimental Conditions (Guardrails).} To systematically evaluate user-facing defenses, we design three guardrails based on a principled, escalating spectrum of intervention intensity: from a passive notice (Guard 1), to a persistent reminder (Guard 2), and finally to an active alert (Guard 3). Our approach draws on established research in usable security warnings~\cite{stock2016hey,zaaba2021harnessing,herley2009so}, which finds that users frequently disregard static notices due to ``warning fatigue.''~\cite{bravo2014harder,stock2018didn} This highlights the need for more dynamic methods of intervention. Our design is further validated by a preliminary survey (see Table~\ref{tab:protection_measures}) in which participants' most requested protections are real-time warnings and periodic reminders, aligning directly with our more active guardrail conditions. As illustrated in Figure~\ref{fig:guardrail}, our three between-subjects conditions are:

\noindent $\bullet$ \textit{Guard 1 (Static Disclaimer)}: A passive, one-time warning shown at the start of the experiment. It mirrors common ``notice and consent'' banners widely used in the privacy compliance domain (e.g., GDPR cookie banners~\cite{degeling2018we}, CCPA notices~\cite{zhang2024cschecker}). Such disclaimers meet the minimal requirements for informing users of potential risks or practices.

\noindent $\bullet$ \textit{Guard 2 (Persistent Reminder)}: A subtle, non-intrusive visual icon and warning that accompanies each agent response. This ambient cue is designed to subtly maintain user awareness of potential risks throughout the interaction, aligning with usable security research that demonstrates the effectiveness of continuous reminders in combating habituation and reinforcing secure behavior~\cite{bravo2014harder,stock2018didn}. Additionally, this design approach is consistent with industry common practices in security-critical applications, such as browser security indicators (e.g., HTTPS lock icons~\cite{ChromeLockIcon2023} and Safe Browsing warnings~\cite{GoogleSafeBrowsing}), which aim to keep users informed without interrupting their primary tasks~\cite{datta2021warning}.

\noindent $\bullet$ \textit{Guard 3 (Interactive Alert)}: An interruptive dialog appears upon suspicious agent output, halting the user’s workflow and requiring explicit acknowledgment. This intervention reflects common practices in the design of high-risk security warnings, which should be context-specific, salient, and actionable to effectively capture user attention and prompt secure decision-making~\cite{blythe2012implementing}. Moreover, such interruptive mechanisms are commonly mandated in critical systems (e.g., operating system security dialogs~\cite{motiee2010windows}) and are increasingly required by AI-specific regulatory frameworks (e.g., NIST AI RMF~\cite{nistrmf2023}; EU AI Act~\cite{euaiact2024}) to ensure meaningful human oversight in high-stakes scenarios.

\smallskip \noindent \textbf{Participant Task Allocation.} To systematically probe human vulnerabilities in agentic systems, we employ a \textit{mixed experimental design}. This approach combines a \textit{between-subjects} design for testing defensive interventions with a \textit{within-subjects} design for assessing vulnerability across various attacks.

\noindent $\bullet$ \textit{Between-Subjects for Guardrails:} To measure the true effect of each defense without confounding learning effects, each participant is randomly assigned to \textit{one and only one} of the three guardrails. Our random assignment successfully results in balanced groups (G1: N=101, G2: N=103, G3: N=99).

\noindent $\bullet$ \textit{Within-Subjects for Attacks:} To assess vulnerability across our diverse scenarios, a key methodological challenge was to manage participant fatigue. Having each participant complete all nine scenarios would require more than 2 hours, an infeasible duration for a crowdsourced study that would introduce significant confounding variables like attention decay and reduced engagement. Thus, we adopt a block design. Each participant is assigned a randomly selected block of three different attack scenarios, keeping the total experiment duration to approximately 40 minutes, an optimal timeframe for high-quality participant engagement. The nine scenarios are grouped into three balanced blocks (details in Appendix~\ref{apx:scenario_blocks}), each containing a mix of attack types and domains.

\begin{table}[t]
    \centering
    \caption{Demographics of study participants \textbf{(N=303)}.}
    \label{tab:demographics}
    \renewcommand{\arraystretch}{0.8}
    \setlength{\tabcolsep}{4pt}
    \resizebox{\linewidth}{!}{
    \begin{threeparttable}
    \begin{tabular}{lrr lrr}
        \toprule
        \textbf{(1) Age} & \textbf{N} & \textbf{\%} & \textbf{(2) LLM Use Freq.} & \textbf{N} & \textbf{\%} \\
        \cmidrule(r){1-3} \cmidrule{4-6}
        \rowcolor{gray!15} 18--24 & 99 & 32.7 & (5) Everyday & 155 & 51.2 \\
        25--34 & 113 & 37.3 & (4) Frequently & 109 & 36.0 \\
        \rowcolor{gray!15} 35--44 & 51 & 16.8 & (3) Occasionally & 29 & 9.6 \\
        45--54 & 20 & 6.6 & (2) Rarely & 6 & 2.0 \\
        \rowcolor{gray!15} 55+    & 20 & 6.6 & (1) Very Rarely & 4 & 1.3 \\
        \midrule
        \textbf{(3) Gender} & \textbf{N} & \textbf{\%} & \textbf{(4) Task-Oriented?} & \textbf{N} & \textbf{\%}  \\
        \cmidrule(r){1-3} \cmidrule{4-6}
        \rowcolor{gray!15} Female & 155 & 51.2 & Yes & 280 & 92.4 \\
        Male   & 148 & 48.8 & No  & 23  & 7.6 \\
        \midrule
        \multicolumn{6}{c}{\textbf{(5) Educational Background}} \\
        \cmidrule(r){1-6}
        \rowcolor{gray!15} Bachelor's degree & 148 & 48.8 & Master's degree & 76 & 25.1 \\
        PhD or above & 46 & 15.2 & High school or below & 33 & 10.9 \\
        \midrule
        \multicolumn{6}{c}{\textbf{(6) Occupation}} \\
        \cmidrule(r){1-6}
        \rowcolor{gray!15} IT \& Technical & 119 & 39.3 & Student & 57 & 18.8 \\
        Business \& Finance & 53 & 17.5 & Education \& Research & 28 & 9.2 \\
        \rowcolor{gray!15} Healthcare \& Medical & 24 & 7.9 & Engineering & 5 & 1.7 \\
        Lifestyle Services & 6 & 2.0 & Others$\natural$ & 11 & 3.6 \\
        \bottomrule
    \end{tabular}%
    \begin{tablenotes}[flushleft]
    \item[] \vspace{-2pt}\hspace{-2pt}$\natural$: Others include law/legal, construction, non-profit organization, transport and warehousing, self-employed, part-time workers, etc.
    \end{tablenotes}
    \end{threeparttable}
    }
\end{table}

\subsection{User Recruitment and Demographics}\label{ssec:method_demograph}
We recruited participants via Prolific~\cite{prolific}, a research crowdsourcing platform known for its high-quality and diverse participant pool. We offered a payment rate of \texteuro9.00 per hour to ensure fair compensation and high engagement. Our eligibility criteria are: (1) Participants are required to be fluent in English, (2) be 18 years or older, (3) use a desktop or laptop computer (mobile devices are disallowed), and (4) have a Prolific submission approval rate of 98\% or higher.
Finally, we collect results from \textbf{N=303} qualified participants after strict data quality checks based on \sys's logging data and users' submitted surveys. The full list of data quality and exclusion rules detailed in Appendix~\ref{apx:data_exclusion}.
As shown in Table~\ref{tab:demographics}, our study involves the core demographic of technically proficient, highly educated power users of today's agentic AI, making their observed vulnerability particularly noteworthy. Our sample is characterized by:

\noindent $\bullet$ \textbf{Deep and Frequent Agentic AI Engagement.} A remarkable 96.7\% of participants used agentic LLMs occasionally, frequently, or every day. The vast majority (93.4\%) use them for specific, goal-oriented tasks, beyond casual exploration. 

\noindent $\bullet$ \textbf{High Educational and Technical Background.} Our participants are highly educated, with nearly 9/10 (89.1\%) holding a bachelor's degree or higher. The largest single group (39.3\%) consists of IT \& Technical professionals, a demographic one might expect to be more sensitive to digital systems.

\noindent $\bullet$ \textbf{Young, Tech-Immersed Group.} The sample is well-distributed across gender (51.2\% female, 48.8\% male) and age (70\% between 18 and 34), a generation has grown up with complex software and AI technology.

It demonstrates that the security gaps (\S\ref{sec:results}) we identify are not simply due to a lack of user sophistication. Rather, they appear to be fundamental flaws in the current human-agent trust model, even affecting those with expertise.

\begin{figure}[t]  
	\centering  
	\includegraphics[width=0.98\linewidth]{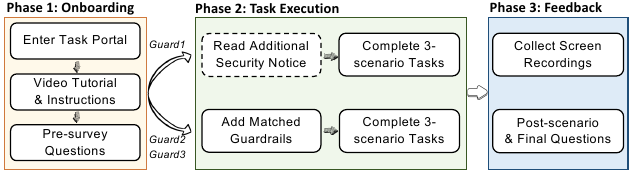}  
	\caption{The three-phase procedure of our experiment. Participants are randomly assigned to a guardrail in Phase 2.}
	\label{fig:userstudy}  
\end{figure}

\subsection{Study Procedure}\label{ssec:method_procedure}
The entire experiment follows a standardized three-phase procedure illustrated in Figure~\ref{fig:userstudy}. The study is designed to take approximately 40 minutes for each participant.

\smallskip \noindent \textbf{Phase 1:} Onboarding and Pre-Survey (5-8 mins). After providing informed consent under our IRB-approved protocol, participants complete a pre-survey to capture their baseline attitudes toward AI systems (e.g., trust, vigilance). They then watch a short tutorial on how to use the \sys platform, which crucially emphasizes a natural, task-oriented mindset to maintain experimental blinding.

\smallskip \noindent \textbf{Phase 2:} Interactive Task Completion (20-25 mins). Participants are assigned to their experimental condition and complete three professional and everyday scenarios. For each scenario, they are given a clear role and objective (e.g., ``You are a project manager; select the most qualified candidate''). They then interact with the dedicated agent instance, using the provided resources (e.g., simulated email inboxes, downloadable documents) to complete their task.

\smallskip \noindent \textbf{Phase 3:} Post-Survey (10-12 mins). After completing all tasks, participants are asked neutral, open-ended questions designed to elicit any sense of anomaly, e.g., ``Did you notice anything unusual or questionable in the agent's output?'' We probe genuine, unprompted user awareness while maintaining experimental blinding.

\subsection{Metrics Used for Analysis}\label{ssec:method_metric}
\smallskip \noindent \textbf{Pre-Survey Metrics (Baseline Attitudes).} To understand pre-existing factors influencing vulnerability, we measure four aspects using a 5-point Likert scale, where higher scores consistently represent a greater degree of the measured trait (e.g., 5 for ``Completely''): (1)~\textit{Trust} in AI systems to complete tasks accurately; (2)~\textit{Vigilance} in double-checking AI outputs; (3)~\textit{Confidence} in identifying potential AI issues; and (4)~\textit{Negative Experience} (a categorical \textit{yes/no/unsure} variable) with prior AI failures. The full questionnaire, including specific anchors for each scale, is detailed in Appendix~\ref{apx:questionnaire}.

\smallskip \noindent \textbf{In-Situ Susceptibility Metrics (During Task).} 
Our primary outcome variables for measuring user susceptibility are: (1) \textit{Risk Perception Rate} (\%): the rate of participants who report noticing anything ``unusual or questionable'' (by answering ``Yes'' or ``Unsure''), capturing an initial sense of anomaly. (2) \textit{Accurate Identification Rate} (\%): the rate of participants who not only perceive a risk but also correctly describe the underlying attack mechanism in a subsequent open-ended question, measuring a deeper understanding of the threat.

\smallskip \noindent \textbf{Post-Survey Metrics (Defense Preferences and Behavioral Impact).} Through the post-survey, we measure the experiment's impact on users, including their (1) \textit{Perceived Severity} of the attack (1-5 Likert scale); (2) \textit{Trust Change} in LLM agentic systems after the study. (3) \textit{Caution Change} in future interactions with agentic systems (\textit{increase/unsure/decrease}). (4) \textit{Preference for Defenses}: users' selection of preferred protective measures, providing direct, user-centered input for the design of safer LLM agentic systems.

\smallskip \noindent \textbf{Statistical Analysis Approach.}
Our analysis relies on a non-parametric statistical framework chosen to match the nature of our data. Given that our study primarily consists of ordinal Likert-scale data and categorical responses, we use the Mann-Whitney U test and Fisher's exact test, which avoid parametric tests' strong assumptions (e.g., normality, interval data).

\begin{figure}[!t]
    \centering
    \includegraphics[width=1\linewidth]{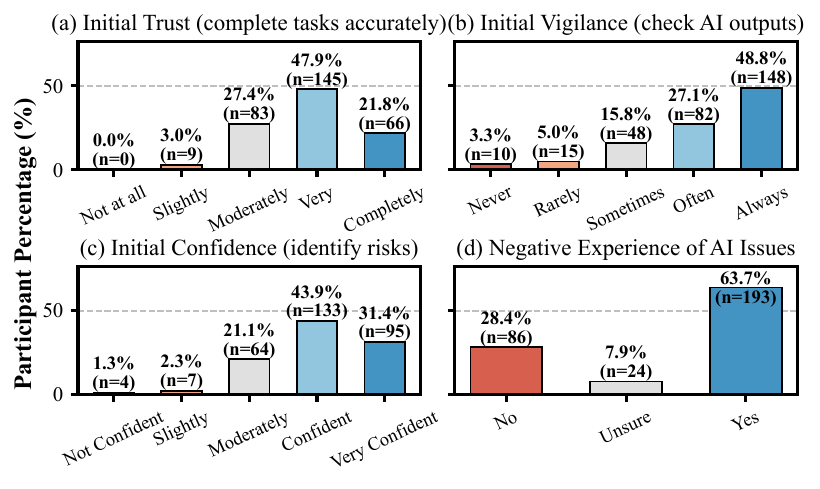}
    \caption{Pre-survey attitudes towards LLM agents (N=303). Users have a mixed, contradictory view. (a) 97.0\% have a moderate to high trust in AI to do tasks correctly. (b) 75.9\% say they check AI outputs often. (c) 96.4\% have a moderate to very high confidence in identifying risks in AI systems. (d) Only 28.4\% report they never experience AI issues.}
    \label{fig:presurvey_3dim}
\end{figure}

\section{Results and Findings}
\label{sec:results}

\subsection{RQ1: How Susceptible Are Users?}\label{ssec:rq1}

Our study characterizes three main aspects of human vulnerability. First, many users have an inaccurate understanding of their own preparedness. Second, we observe that experts could be more susceptible than beginners. Third, we see a gap between risk awareness and protective behavior.

\smallskip \noindent $\bullet$ \textbf{Finding 1: Users think they are prepared, but a mix of high trust, superficial vigilance, and overconfidence makes them susceptible.}
Our pre-survey shows that users begin with a risky mindset. For example, 97.0\% of users trust AI agents to perform tasks correctly, and 75.9\% say they check the AI's work (Figure~\ref{fig:presurvey_3dim}). But this checking is not very deep. One user says he only looks for small mistakes: \textit{``I trust the AI to get the logic right, I just check for simple syntax errors.''}
On top of this, users are very overconfident, which is also documented in phishing~\cite{wang2016overconfidence}. About 75.3\% feel sure they can perceive problems with AI. We call this an ``\textbf{Illusion of Preparedness.}'' The combination of high trust, shallow checking, and overconfidence increases human susceptibility from the start.
Also, Table~\ref{tab:AIexperience_3dim} shows that even users who have negative AI experiences are still just as overconfident (3.96, p = 0.024). They become more cautious (4.31, p = .0003), but remain equally confident in their own skills.

\begin{table}[t]
    \centering
    \caption{Effect of negative AI experience on user attitudes.}
    \label{tab:AIexperience_3dim}
    \renewcommand{\arraystretch}{0.9}
    \setlength{\tabcolsep}{2pt}
    \resizebox{\linewidth}{!}{
    \begin{threeparttable}
    \begin{tabular}{
        l 
        c 
        c 
        c
        c    }
    \toprule
    \multirow{2}{*}{\textbf{Metric}} & \textbf{Experienced} & \textbf{Inexperienced} & {\textbf{Change}} & \multirow{2}{*}{\textbf{\textit{p}-value}\tnote{$\star$}} \\
    & \textbf{(n=193)} & \textbf{(n=110)\tnote{$\natural$}} & {\textbf{(Abs. (Rel. \%))}} & \\
    \midrule
    Trust (1--5) $\downarrow$     & \textbf{3.65\(\pm\)0.70} & 4.30\(\pm\)0.72 & \textbf{-0.65(-15.1\%)}  & \sig{$<$.001} \\
    Vigilance (1--5) $\uparrow$    & \textbf{4.31\(\pm\)0.92} & 3.82\(\pm\)1.21 & \textbf{+0.49 (+12.8\%)}  & \sig{.0003} \\
    Confidence (1--5) $\downarrow$ & \textbf{3.96\(\pm\)0.77} & 4.11\(\pm\)0.99 & \textbf{-0.15 (-3.6\%)}  & \sig{0.024} \\
    \bottomrule
    \end{tabular}
    \begin{tablenotes}[flushleft]
        \item[] \vspace{-2pt}\hspace{-2pt}$\natural$: $p$-values are from Mann-Whitney U test. \sig{Red} denotes statistical significance.
    \end{tablenotes}
    \end{threeparttable}
    }
\end{table}

\smallskip \noindent $\bullet$ \textbf{Finding 2: Users fail due to task-focused tunneling, with experts facing an additional domain knowledge trap.}
The ``Illusion of Preparedness'' from Finding 1 breaks when users face real attacks. The failure rate is high. In our basic test, only 8.6\% of users perceive risks. This shows a big gap in how users understand agent-mediated deception. We find that all users suffer from the same fundamental cognitive error: task-focused tunneling that leaves no mental capacity for security vigilance.
As one user explains, \textit{``It's a user-friendly platform. I didn't notice any risk alerts as I was more focused on solving problems.''} Users become cognitively absorbed in task completion, leaving no attention for threat perception. This explains why only 24\% of users even notice the initial security disclaimer (i.e., Guard 1) (Figure~\ref{fig:guardrail_overall}).
However, domain professionals\footnote{Domain professionals are participants with advanced IT/coding expertise. General users are highly educated and competent (with bachelor's degrees or higher), but lack the same depth of specialized experience. This distinction is about domain-specific expertise, not general cognitive ability.} face an additional vulnerability we term the ``Expert's Paradox.'' Domain professionals consistently underperform general users across all conditions (e.g., 6.8\% vs. 9.6\% in G1). This is most dramatic in code-related scenarios where IT/technical professionals achieve 0.0\% and 9.1\% perception while general users reach 8.1\% and 13.5\%, respectively (Table~\ref{tab:perception_technical}). This paradox stems from experts applying their domain knowledge to evaluate \textit{what} the agent produces (content quality) while unconsciously delegating trust in \textit{how} the agent operates (process integrity). Their specialized knowledge becomes a cognitive tunnel that not only blocks peripheral threat perception but actively reinforces trust in the agent's competence. As Table~\ref{tab:attitude_4dim} shows, users who successfully perceive attacks begin with significantly lower baseline trust (3.68 vs. 3.96, p=0.021), suggesting that expertise-driven confidence systematically impairs threat sensitivity.

\begin{table}[t]
    \centering
    \caption{Risk perception rates for users with different technical expertise. Domain experts show lower perception rates.}
    \label{tab:perception_technical}
    \renewcommand{\arraystretch}{0.95}
    \setlength{\tabcolsep}{8pt} 

    \resizebox{\linewidth}{!}{
    \begin{threeparttable}        
    \begin{tabular}{@{}l ccc@{}}
        \toprule
        \textbf{Group / Perception Rate (\%)} & \textbf{Guard 1} & \textbf{Guard 2} & \textbf{Guard 3} \\
        \midrule
        \quad Overall           & 8.6\%\textsuperscript{[183]}  & 20.0\%\textsuperscript{[60]} & 25.0\%\textsuperscript{[60]} \\
        \quad Domain Professionals        & 6.8\%\textsuperscript{[69]}  & 18.4\%\textsuperscript{[29]} & 22.2\%\textsuperscript{[21]} \\
        \quad General Users       & \textbf{9.6\%\textsuperscript{[114]}}  & \textbf{21.5\%\textsuperscript{[31]}} & \textbf{26.5\%\textsuperscript{[39]}} \\
        \bottomrule
    \end{tabular}
    \begin{tablenotes}[flushleft]
        \item[] \hspace{-2pt}(1) CodeRepoAgent: Domain Prof. (0.0\%) vs General Users (\textbf{8.1\%}). 
        \item[] \hspace{-2pt}(2) CodeEvalAgent: Domain Prof. (9.1\%) vs General Users (\textbf{13.5\%}).
    \end{tablenotes}
    \end{threeparttable}
    }
\end{table}

\begin{figure}[t]
    \centering
    \includegraphics[width=0.9\linewidth]{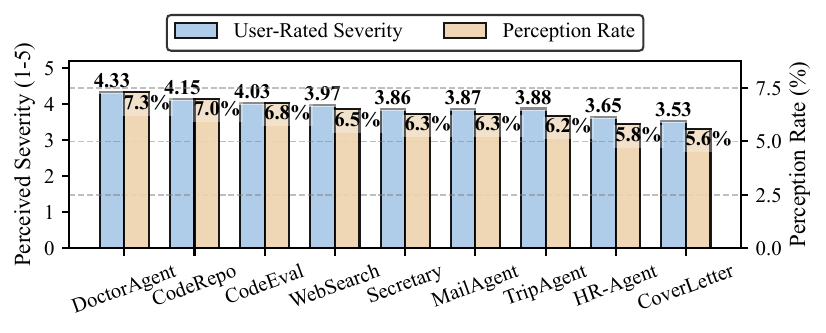}
    \caption{User-rated severity and attack perception rate across nine agent scenarios. A positive correlation is observed, where higher severity is associated with higher perception rate.}
    \label{fig:perception_per_scenario}
\end{figure}

\smallskip \noindent $\bullet$ \textbf{Finding 3: Users know when a task is risky, but they don't act any safer.}
Users are good at telling which tasks are risky. Figure~\ref{fig:perception_per_scenario} shows they know the DoctorAgent task is the most dangerous, rating its severity at 4.33 out of 5. But this awareness does not change how they acted. We call this the ``Awareness-Action Gap.''
The results show this gap clearly. Even in the medical task that they rate as most risky, users only perceive 7.3\% of the attacks. This means over 92\% of attacks still got through. Their general knowledge about the risk does not help them when they are in the middle of the task.
This finding shows that knowing a risk and acting on it are two different things. A user is aware that errors in a task could lead to serious consequences, but still fall into the same cognitive traps we see in Finding 2. This means that just teaching users about risks is not enough to protect them.

\boxx{\textbf{RQ1 Takeaway:} Human vulnerability to AMD is not an anomaly but a structural feature of the current human-agent trust paradigm. Factors that typically signify competence (e.g., domain expertise and deep task focus) are paradoxically transformed into the most effective attack vectors, undermining user judgment.}

\subsection{RQ2: What Factors Impact Susceptibility?} 
\label{ssec:perception_analysis}
Having established the scale of user vulnerability, we now examine the specific psychological patterns and individual differences that predict susceptibility. Our findings identify distinct cognitive failure modes and reveal that resilience stems from mindset traits rather than technical competence.

\smallskip \noindent $\bullet$ \textbf{Finding 4: We find six common thinking patterns that make users susceptible.}
Our analysis finds that user failures are caused by six common thinking patterns.

\textit{\ding{202} Task-Focused Tunneling.} Users get very focused on their main task, leaving no attention for security. One IT professional says, \textit{``I didn't notice any risk alerts as I was more focused on solving problems.''} That is why only 24\% of users (Figure~\ref{fig:guardrail_overall}) confirm the Guard 1 security notice at the start.

\textit{\ding{203} Partial Verification.} Users check the first few things an agent does. If they look right, they trust everything else after that. One student user says, \textit{``I only checked the first few things if they matched... if they did I trusted the bot.''}

\textit{\ding{204} Utility Override.} If an agent is very useful, users will ignore security risks to get the benefit. One IT professional says, \textit{``Despite security concerns, I saw how AI agents can be highly effective at reducing manual workload.''}

\textit{\ding{205} Transparency Preference.} Security warnings can make users trust a system more, not less. One expert user says a warning makes them feel \textit{``more informed and cautious, which improved my overall sense of safety.''} They see the warning as a sign of a good system.

\textit{\ding{206} Algorithmic Adherence.} Users assume that if an agent performs well, it should also be safe. They hand over their judgment to the algorithm. This is a special problem for users with IT/technical expertise, who are good at judging performance but then wrongly equate it with security.

\textit{\ding{207} Vigilance Degradation.} Good experiences make users less careful over time, which opens the door for risks. A teacher user says, \textit{``I used to think that AI needed to be double/triple checked after use, but this was just fine.''}

\textbf{The Cognitive Escalation Effect.} These patterns do not occur in isolation; they reinforce each other in a predictable sequence that leads to complete security breakdown. The process typically begins with Utility Override: users discover the agent's remarkable effectiveness. As one user notes, \textit{``I have noticed how many types of tasks I can do with it in a very short period of time.''} This positive experience triggers Partial Verification (checking less while experiencing continued success), which amplifies trust. This creates a downward spiral culminating in complete Algorithmic Adherence, where users surrender their judgment entirely: \textit{``AI is effective and doesn't rely on a human being... just your information and the rest AI will do for you.''}

\smallskip \noindent $\bullet$ \textbf{Finding 5: Resilience is a ``security mindset'' and active skepticism to question AI agent's behavior.}
Our analysis pinpoints the crucial factor differentiating resilient users from susceptible ones: a pre-existing cognitive disposition we term the ``Security Mindset.'' This mindset is not about technical knowledge, but a default way of thinking characterized by healthy skepticism.
This mindset is characterized by a default posture of healthy skepticism. As shown in Table~\ref{tab:attitude_4dim}, users who successfully perceive an attack (the ``Perceived'' group) enter the study with a distinct psychological profile. Compared to those who do not, they exhibit significantly lower initial Trust (3.68 vs. 3.96, p = .021), higher self-reported Vigilance (4.31 vs. 4.06, p = .045), and were far more likely to have had prior negative Experience with AI systems (p $<$ .001). This confirms that a skeptical default is often forged through past negative experiences.
\begin{table}[t]
    \centering
    \caption{Attitudes of users who perceive an attack vs. those who do not. Perceived users start with lower trust, higher caution, and are highly likely to have experienced AI errors.}
    \label{tab:attitude_4dim}
    \renewcommand{\arraystretch}{0.9}
    \setlength{\tabcolsep}{2pt}

    \resizebox{\linewidth}{!}{
    \begin{threeparttable}
    \begin{tabular}{
        l 
        c 
        c 
        c
        c
    }
    \toprule
    \multirow{2}{*}{\textbf{Metric}} & \textbf{Perceived} & \textbf{Unperceived} & {\textbf{Change}} & \multirow{2}{*}{\textbf{\textit{p}-value}\tnote{$\star$}} \\
    & \textbf{(n=85)} & \textbf{(n=218)} & {\textbf{(Abs. (Rel. \%))}} & \\
    \midrule
    Trust (1--5) $\downarrow$      & \textbf{3.68\(\pm\)0.76} & 3.96\(\pm\)0.77 & \textbf{-0.28 (-7.1\%)}  & \sig{0.007} \\
    Vigilance (1--5) $\uparrow$    & \textbf{4.31\(\pm\)0.98} & 4.06\(\pm\)1.09 & \textbf{+0.24 (+5.9\%)}  & \sig{0.054} \\
    Confidence (1--5) $\downarrow$ & 3.98\(\pm\)0.80 & 4.03\(\pm\)0.88 & -0.056 (-1.4\%)  & 0.392 \\
    Experience\tnote{$\natural$}~~$\uparrow$  & \textbf{0.81\(\pm\)0.39} & 0.65\(\pm\)0.48 & \textbf{+0.16 (+25.6\%)} & \sig{0.007} \\
    \bottomrule
    \end{tabular}
    \begin{tablenotes}[flushleft]
    \item[] \vspace{-2pt}\hspace{-2pt}$\natural$: Prior negative experience is in binary scale: Yes=1, No=0.
    $\star$: $p$-values are from Mann-Whitney U test. \sig{Red} denotes statistical significance.
    \end{tablenotes}
    \end{threeparttable}
    }
\end{table}

Furthermore, this ``Security Mindset'' is not a passive trait but an active, behavioral practice of proactive scrutiny. We see this clearly in Table~\ref{tab:proactive_ask}, which shows that 37.0\% of users who report actively relying on their ``personal judgment'' to question the AI's output are dramatically more successful at perceiving risks. Their overall risk perception rate is 19.6\%, a +39.5\% lift compared to the baseline rate. This effect is most pronounced with the weakest defenses, where proactive users in the Guard 1 condition achieve a +51.2\% lift in perception. This demonstrates that resilience is an active decision to challenge and verify information, rather than to passively accept the output of a trusted agent.

\begin{table}[!t]
    \centering
    \caption{How proactive user judgment affects risk perception. Users exhibiting proactive scrutiny (relying on personal judgment) show higher risk perception rates.}
    \label{tab:proactive_ask}
    \renewcommand{\arraystretch}{0.8}
    \setlength{\tabcolsep}{4pt}
    \resizebox{\linewidth}{!}{
    \begin{threeparttable}
    \begin{tabular}{l
            r
            S[table-format=2.1]
            S[table-format=2.1]
            S[table-format=2.1]
            S[table-format=2.1, round-precision=1, table-sign-mantissa, table-format = +2.1]
            S[table-format=2.1]}
        \toprule
        \multirow{2.5}{*}{\textbf{Guardrail}} & {\textbf{Num}} & {\textbf{Prop.}} & \multicolumn{2}{c}{\textbf{Risk Perc. (\%)}} & {\textbf{Perc.}} & {\textbf{Prop. in}} \\
        \cmidrule(lr){4-5}
        & {\textbf{Users}} & {\textbf{Proact.}} & {\textbf{All}} & {\textbf{Proact.}} & {\textbf{Lift}} & {\textbf{Perc. Users}} \\
        \midrule
        \rowcolor{gray!15}
        G1 (Static)      & 183 & 40.8 & 8.6  & \textbf{12.9} & \textbf{+51.2}  & 61.7 \\
        G2 (Persistent)  & 60  & 26.1 & 20.0 & \textbf{34.0} & \textbf{+70.2}  & 44.4 \\
        \rowcolor{gray!15}
        G3 (Interactive) & 60  & 36.1 & 25.0 & \textbf{32.3} & \textbf{+29.2}  & 46.7 \\
        \midrule
        \textbf{Overall} & \textbf{303} & \textbf{37.0} & \textbf{14.1} & \textbf{19.6} & \textbf{+39.5} & \textbf{51.6} \\
        \bottomrule
    \end{tabular}
    \begin{tablenotes}[flushleft]
        \item[] \vspace{-2pt}\hspace{-2pt}$\natural$: Prop. Proact. = proportion of users who reported relying on personal judgment; Risk Perc. All/Proact. = risk perception rate for all users/proactive users only; Perc. Lift = percentage point improvement for proactive users; Prop. in Perc. Users = proportion of proactive users among those who successfully perceived attacks.
    \end{tablenotes}
    \end{threeparttable}  
    }
\end{table}

\smallskip \noindent $\bullet$ \textbf{Finding 6: Multi-factor analysis confirms that how users think is more important for security.}
To identify the cognitive factors that drive susceptibility to AMD attacks, we conduct a multi-factor analysis, focusing on how user traits influence threat perception (Table~\ref{tab:multifactor}). 
Our findings are organized into three tiers based on statistical significance.

\textit{Tier 1: Key Predictors (\textit{p}$<$.0001).} Two cognitive factors are predictive of susceptibility: Consistency (judging whether the agent-generated content is aligned with expectation) and In-Situ Trust (trust in AI's correctness during tasks). Users who identify inconsistencies were nearly four times more likely to perceive attacks (24.29\% vs. 6.91\%, p$<$.0001), supporting the idea of a ``Security Mindset'' (an active, questioning approach). Similarly, low-trust users were more successful in identifying threats than high-trust ones (23.19\% vs. 7.02\%, p$<$.0001). These findings underscore that vulnerability is shaped by real-time cognitive processes, not inherent traits.

\textit{Tier 2: Moderate Predictors (\textit{p}$<$.05).} Some traditional markers of competence, such as Education and Age, have weaker or unexpected effects. Higher education (M.S./Ph.D.) correlates with worse performance compared to basic education (4.64\% vs. 10.68\%, p=.0028), demonstrating the ``Expert's Paradox'' (Finding 2). Younger users (18-34) outperform older users (9.43\% vs. 5.49\%, p=.0307), suggesting digital familiarity is a stronger protective factor than experience. Negative AI experience provide modest benefits (9.52\% vs. 5.04\%, p=.0112), indicating that experiential learning often outperforms formal education.

\begin{table}[!t]
\centering
\caption{Multi-factor correlation analysis of risk perception. Results reveal that cognitive processes during interaction are more predictive than pre-existing user characteristics.}
\label{tab:multifactor}
\resizebox{\linewidth}{!}{
\begin{threeparttable}
\renewcommand{\arraystretch}{0.85}
\setlength{\tabcolsep}{12pt}
\begin{tabular}{@{}llcccccccc@{}}
\toprule
\multirow{2.5}{*}{\textbf{Factor}} & \multirow{2.5}{*}{\textbf{Group}} & \multicolumn{2}{c}{\textbf{Perception Rate}} \\
\cmidrule(lr){3-4} 
& & (\%) $\uparrow$ & \textit{p}-value\tnote{$\star$} \\
\midrule

\multirow{2}{*}{\textbf{Consistency}} & Low ($\le$2) & \textbf{24.29} & \multirow{2}{*}{\sig{$<$.0001}} \\
 & High ($\ge$3) & 6.91 & \\
\midrule

\multirow{2}{*}{\textbf{In-Situ Trust}} & Low ($\le$2) & \textbf{23.19} & \multirow{2}{*}{\sig{$<$.0001}} \\
& High ($\ge$3) & 7.02 & \\
\midrule

\multirow{2}{*}{\textbf{In-Situ Stress}} & Low ($\le$2) & \textbf{8.35} & \multirow{2}{*}{.1645} \\
 & High ($\ge$3) & 8.15 & \\
\midrule

\multirow{2}{*}{\textbf{Education}} & B.S., H.S. or below & \textbf{10.68} & \multirow{2}{*}{\sig{.0028}} \\
& M.S., Ph.D. or above & 4.64 & \\
\midrule

\multirow{2}{*}{\textbf{Age}} & 18-34 & \textbf{9.43} & \multirow{2}{*}{\sig{.0307}} \\
& $\ge$35 & 5.49 & \\
\midrule

\multirow{2}{*}{\textbf{AI Use Frequency}} & Low ($\le$3) & \textbf{10.26} & \multirow{2}{*}{.209} \\
& High ($\ge$4) & 7.95 & \\
\midrule

\multirow{2}{*}{\textbf{Negative Exp.}} & Yes / Unsure & \textbf{9.52} & \multirow{2}{*}{\sig{.0112}} \\
& No & 5.04 & \\
\midrule

\multirow{2}{*}{\textbf{Init Trust}} & Low ($\le$3) & 8.21 & \multirow{2}{*}{.3467} \\
& High ($\ge$4) & \textbf{8.33} & \\
\midrule

\multirow{2}{*}{\textbf{Init Vigilance}} & Low Frequency ($\le$3) & 8.22 & \multirow{2}{*}{.7938} \\
& High Frequency ($\ge$4) & \textbf{8.26} & \\
\midrule

\multirow{2}{*}{\textbf{Init Confidence}} & Low ($\le$3) & \textbf{8.89} & \multirow{2}{*}{.9157} \\
& High ($\ge$4) & 8.04 & \\

\bottomrule
\end{tabular}
\begin{tablenotes}[flushleft]
    \item[] \vspace{-2pt}\hspace{-2pt}$\star$: \textit{p}-values are from Fisher's exact test. \sig{Red} denotes statistical significance.
\end{tablenotes}
\end{threeparttable}
}
\end{table}

\textit{Tier 3: No Predictive Value, Confirming the ``Illusion of Preparedness''.} Factors such as Initial Confidence, Initial Trust, and Initial Vigilance have no predictive value. %
Specifically, users' self-assessed abilities do not correlate with their attack perception rate. Such perception for users with high self-confidence versus those with low self-confidence is statistically indistinguishable (8.89\% vs. 8.04\%, p=.9157). 
This null result reveals a misalignment between users' perceived readiness and their actual capabilities, which is the very definition of the ``Illusion of Preparedness.'' (see our Finding 1 in \S\ref{ssec:rq1}), that users operate under the false belief that their confidence or caution is a protective asset, when in reality it has no bearing on their likelihood of perceiving a threat.

\boxx{\textbf{RQ2 Takeaway:} Susceptibility to AMD is not determined by static user traits like expertise or confidence. Instead, it follows predictable cognitive failure paths, where users escalate from trust to blind acceptance. A ``security mindset'' is more effective, which actively questions the agent's process, not just its final output.}

\subsection{RQ3: How To Design Better Defenses?}
\label{ssec:defense_evaluation}
We explore building defenses that actually work: ``How to design better defenses against these flawed thinking patterns?''

\begin{figure}[t]
    \centering
    \includegraphics[width=1\linewidth]{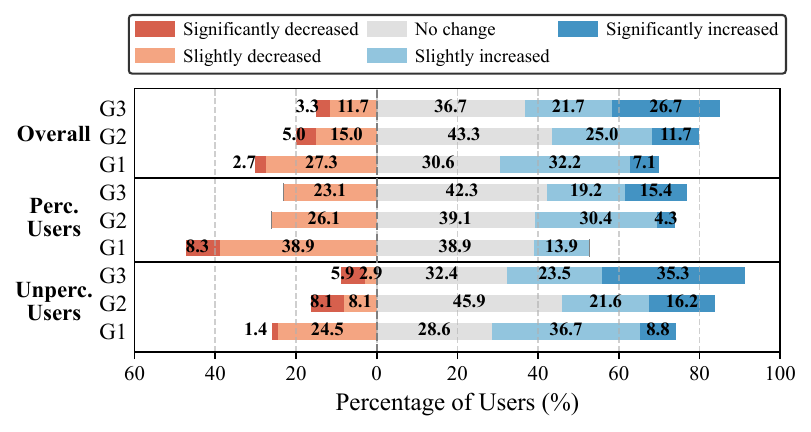}
    \caption{Changes in user trust after interacting with \sys, broken down by guardrail (G1, G2, G3) and whether the user perceives a risk. Users who missed an attack often reported increased trust, which makes them more susceptible.}
    \label{fig:trust_change}
\end{figure}

\smallskip \noindent $\bullet$ \textbf{Finding 7: Users want control and transparency, but they should be cautious with the transparency-trust paradox.}
We ask users what would make them feel safer using agentic systems. Table~\ref{tab:protection_measures} shows they want more transparency and user control: ``real-time warnings'' (61.1\%), and ``periodic reminders'' (55.8\%). Users explicitly want visibility into system operations rather than black-box automation. These features exactly match our designed (G3) interactive alert and (G2) persistent reminder. However, Figure~\ref{fig:trust_change} reveals a troubling paradox: users who do not perceive issues but experience G2 and G3 often report more significantly increased trust. %
An IT professional (unperceived user) explains \textit{``the security/risk alert made me more aware of potential threats... which improved my overall sense of safety while using \sys.''} Another participant notes: \textit{``The presence of alerts and transparency features improved my confidence in these tools when properly monitored.''} A third describes how \textit{``the system demonstrated consistency, transparency in evaluation, and highlighted even subtle issues, showing thoughtful safeguards are in place.''} This directly validates the ``Transparency Preference'' pattern from Finding 4: users interpret security warnings as indicators of system trustworthiness rather than markers of risk.

\begin{figure}[t]
    \centering
    \includegraphics[width=0.7\linewidth]{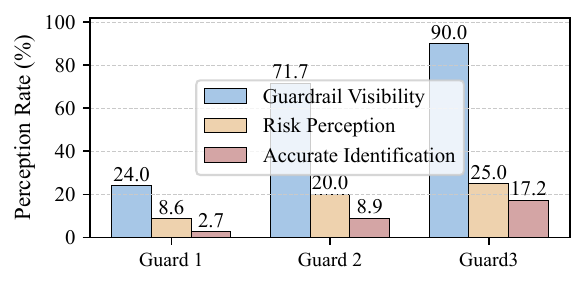}
    \caption{Effectiveness of the three guardrails. A clear order of effectiveness shows: G3 $>$ G2 $>$ G1, from (i) guardrail visibility, to (ii) risk perception, to (iii) accurate identification.}
    \label{fig:guardrail_overall}
\end{figure}

\begin{figure}[t]
    \centering
    \includegraphics[width=1\linewidth]{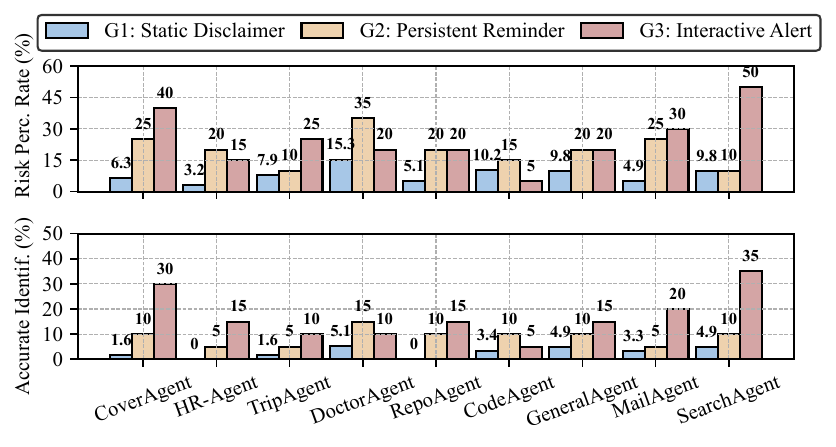}
    \caption{Per-scenario breakdown of guardrail effectiveness. Guard 2 and Guard 3 consistently yield the highest rates of both risk perception (top) and accurate identification (bottom).}
    \label{fig:guardrail_scene}
\end{figure}

\smallskip \noindent \textbf{Finding 8: Effective warnings should interrupt users, while featuring low verification cost.}
Our study explains that most security warnings fail due to two challenges. First, they should break the user's task focus. Second, they should allow users to verify risks without heavy cognitive costs.
Figure~\ref{fig:guardrail_overall} shows that interruption is essential. Static disclaimer (G1) has a 2.7\% perception rate, while interactive alerts (G3) reach 17.2\%. However, Figure~\ref{fig:guardrail_scene} reveals that interruption alone is not enough. The effectiveness of guardrails varies widely across scenarios.
For tasks like CoverAgent and SearchAgent, G3 leads to large gains in risk perception (from 6.3\% to 40\%, and from 9.8\% to 50\%, respectively). In contrast, complex tasks like TripAgent and MailAgent show much smaller improvements (TripAgent: 7.9\% to 15\%; MailAgent: 9.3\% to 15\%).
This difference reflects the cognitive cost of verification. When verifying is easy (e.g., checking a letter format or search result), users benefit from the interruption. But in tasks requiring detailed checks (e.g., reviewing travel itineraries or email authenticity), users face cognitive overload. Even when interrupted, they struggle to process the information fully.
Moreover, Figure~\ref{fig:guardrail_scene} (bottom) shows that accurate threat identification remains low across all scenarios. Many users perceive something is wrong but cannot specify the threat. This suggests that while interruptions can raise awareness, they are insufficient when verification demands exceed users' mental capacity. Users fall back on guessing rather than systematic analysis.

\smallskip \noindent $\bullet$ \textbf{Finding 9: Effective defense creates necessary friction, turning usability costs into resilient trust.}
Our study shows a nuance in the usability-security trade-off. While engaging with a threat incurs interruptive alerts that break users' workflow, this friction is not a simple usability penalty. As shown in Table~\ref{tab:guard_pleasant_recommend}, users in the most effective defense (G3) report the highest overall pleasantness (3.96 vs. 4.59, p=0.007) and recommendation ratings (4.66 vs. 4.45, p=0.205). This indicates that users view the interruption not as mere annoyance, but rather as a valuable and necessary safeguard from a system that is actively protecting them. 
This positive perception is directly linked to the cultivation of earned trust. Our data show that increased trust consistently translates into a higher willingness to recommend the system (e.g., 4.43 vs. 4.03, p=.002 in G1).
Therefore, designers should not aim to eliminate all friction, but to design meaningful friction, i.e., interruptions that serve as a clear investment in building a resilient, long-term trust relationship with the user. As one user says, the alert \textit{``improved my overall sense of safety.''} 
\begin{table}[t]
    \centering
    \caption{Pleasantness ratings and trust effects on recommendation likelihood. Left: How \sys's pleasantness ratings differ between users who perceived an attack and those who did not. Right: How trust affects recommendation likelihood.}
    \label{tab:guard_pleasant_recommend}
    \renewcommand{\arraystretch}{0.9}
    \setlength{\tabcolsep}{2pt}
    \resizebox{\linewidth}{!}{
    \begin{threeparttable}
    \begin{tabular}{
        l                 
        c c c            
        c c c            
    }
    \toprule
    \multirow{2.5}{*}{\textbf{Condition}} 
    & \multicolumn{3}{c}{\textbf{Pleasantness} $\uparrow$} 
    & \multicolumn{3}{c}{\textbf{Recommendation} $\uparrow$} \\
    
    \cmidrule(lr){2-4} \cmidrule(lr){5-7}
    
    & {Perceived} & {Unperceived} & {\textit{p}-value} 
    & {Trust+\tnote{$\natural$}} & {Trust--} & {\textit{p}-value} \\
    
    \midrule
    
    Guard 1 
    & 3.86\(\pm\)0.83 & 4.12\(\pm\)0.97 & 0.117
    & 4.43\(\pm\)0.65 & 4.03\(\pm\)1.10 & \sig{0.002} \\

    \addlinespace[2pt]
    
    Guard 2
    & 3.87\(\pm\)1.10 & 4.30\(\pm\)0.81 & 0.115
    & 4.50\(\pm\)0.74 & 4.16\(\pm\)0.86 & 0.110 \\

    \addlinespace[2pt]

    Guard 3
    & \textbf{3.96\(\pm\)0.87} & \textbf{4.59\(\pm\)0.86} & \sig{0.007}
    & \textbf{4.66\(\pm\)0.55} & \textbf{4.45\(\pm\)0.68} & 0.205 \\

    \bottomrule
    \end{tabular}
    
    \begin{tablenotes}[flushleft]
        \item[] \vspace{-2pt}\hspace{-2pt}$\natural$`Trust+' refers to users whose trust increased after interaction; `Trust--' refers to users whose trust decreased or remained unchanged.
    \end{tablenotes}
    \end{threeparttable}
    }
\end{table}

\begin{figure}[t]
    \centering
    \includegraphics[width=1\linewidth]{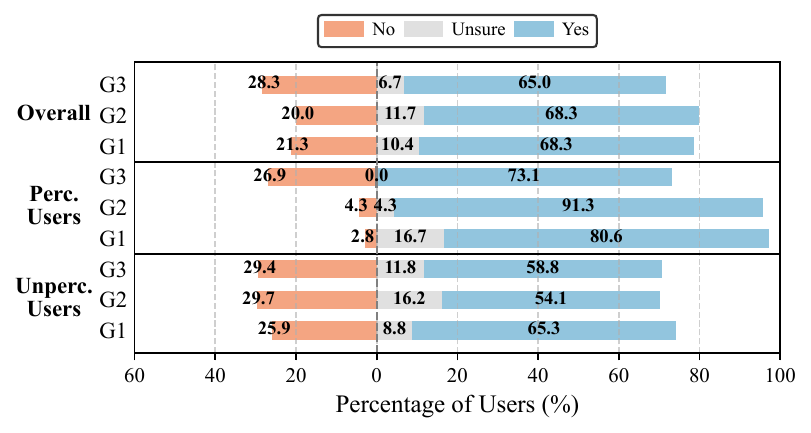}
    \caption{User responses to ``\textit{Will you be more cautious in the future?}'' The experience with \sys is a powerful lesson, especially for users who perceived AMD attacks.}
    \label{fig:cautious_change}
\end{figure}

\smallskip \noindent $\bullet$ \textbf{Finding 10: An effective way to educate users is through experiential learning with simulated attacks.}
Our findings suggest that direct experience is a powerful teacher. Letting users face a simulated threat can be more effective at changing their security mindset than abstract warnings about risks.
The effect on behavior is clear. Figure~\ref{fig:cautious_change} shows that users who actually perceive an attack are much more likely to say they will be more cautious in the future (over 90\% in G2 and G3). The personal experience of finding a threat changes behavior in a way that abstract warnings cannot.
The effect on thinking is even deeper. The experience turns users' vague fears into specific threat models. General worries about ``Privacy Risks'' (49.8\%) become concrete concerns about ``data leakage during inter-tool transfers.'' Such findings suggest that experience makes users not only more cautious, but also smarter in their approach to security.
This points to a new way to think about security training. It suggests that improving user safety is less about refining warnings and more about providing seamless user experiences. Our study shows the power of a ``security flight simulator'': a safe place for users to learn from failure. This helps them build the thinking patterns they need to stay safe in the real world.

\boxx{\textbf{RQ3 Takeaway:} The future of usable security for LLM agents lies not in eliminating all barriers, but in introducing calibrated friction. Effective defense should be conceived as an interactive process that fosters user resilience, rather than a static mechanism that attempts to shield passive users.}
\section{Discussion}\label{sec:discussion}
\noindent \textbf{Rethinking agentic threat models: The human as the last line of defense.}
Our work acknowledges the value of agent-centric security but contends that exclusive reliance on technical defenses is insufficient and risky. By design, agentic systems operate in open and often untrusted environments, where no technical safeguard can be considered fully reliable. In such settings, the human user frequently becomes the implicit final line of defense. Therefore, the central question should evolve from ``Can the agent be broken?'' to ``What happens when an agent is weaponized against its user?'' Our trust boundary framework provides a structured way to analyze these threats, focusing specifically on how the unique capabilities of agents can be subverted to deceive the user. This is distinct from classical LLM safety, security, and privacy; it targets the unique components of agents that LLMs do not have (i.e., \textit{Perception}, \textit{Memory}, and \textit{Action}).

\smallskip \noindent \textbf{Implications for defensive design.} Our findings suggest that effective defenses for agent-mediated deception require a new philosophy: moving away from the traditional goal of a frictionless user experience. Instead, we advocate for carefully engineered interventions that prompt user scrutiny without overwhelming them. Specifically, results show these interventions should first be interruptive to break through task-focused tunneling (Finding 4). Second, they need to present threats with low verification costs to avoid cognitive overload. Third, they should provide actionable provenance to ground user trust in verifiable evidence rather than blind faith (Finding 9). This combination of design choices forms a robust strategy to empower users and build resilience. %

\smallskip \noindent \textbf{Implications for security education.} Our findings signal a new security education paradigm in the agentic era, i.e., a shift in user posture from passive suspicion, as in phishing, to proactive task delegation to agents. Simply telling users to ``be careful'' is ineffective when their entire mental model is primed for collaboration, not skepticism. This is why abstract, one-time warnings (Guard 1) failed so completely in our study (Figure~\ref{fig:guardrail_overall}).
To address this gap, we should move beyond teaching rules to fostering resilient mental models. This requires a new form of training: immersive, experiential education. Our work demonstrates that direct experience with a safe failure can be an effective teacher, significantly shifting users' future caution and threat awareness (Finding 10, Figure~\ref{fig:cautious_change}).
This points to the necessity of building ``security flight simulators'' for LLM agents. \sys is not just a research platform, but an initial prototype of this simulator concept. Its goal is to provide users with a safe environment to experience realistic agent-mediated deception, learn from their mistakes, and thereby build the robust, intuitive mental models necessary to navigate the complexities of human-agent interaction security.

\smallskip \noindent \textbf{Limitations and future work.} Our large-scale study, while systematic, has limitations that open avenues for future research. (1) Our cross-sectional study provides an important snapshot of user vulnerability. However, understanding how these trust dynamics and vulnerabilities evolve requires longitudinal research. Our \sys platform serves as a good foundation to conduct in-situ studies within organizations, tracking user behavior over weeks or months. This would allow for an investigation of complex phenomena such as vigilance decay and the long-term effects of security training. 
(2) While our discovery of the Expert’s Paradox is a key finding, our experts are primarily defined by their IT/technical background. Whether this paradox holds for experts in other high-stakes domains, such as medicine, human resources, or finance, remains an open question. This is our future direction to understand their unique cognitive biases and trust patterns.
(3) Although we consider diversified, realistic everyday and professional attacks, our attacks are statically configured. A future direction is to explore adaptive attacks that learn and exploit a specific user's pattern of trust.

\section{Conclusion}
This work presents a large-scale empirical study (N=303) of human vulnerability to AMD threats. Our findings reveal a landscape of significant and counter-intuitive failures: only 2.7\% of users correctly identify AMD, while domain experts paradoxically prove more susceptible. We identify six distinct cognitive failure modes, showing these issues stem from flawed mental models, not user carelessness. 
Our findings demand a paradigm shift in AI security, expanding the focus from agent-centric defenses to empowering the human user as the last line of defense. This requires a new design philosophy of calibrated friction and a new security education built on experiential learning in ``security flight simulators''.

\section*{Ethical Considerations}
The study was approved by our Institutional Review Board (IRB). All participants provided informed consent before starting the study; they were advised of their right to withdraw at any time without penalty. We provided fair compensation based on Prolific's recommended payment standards for participants' time and effort.
No personally identifiable information (PII) was collected. While we recorded basic demographic data (e.g., age, education, prior LLM experience) and in-study interactions, this information was linked only to anonymized Prolific IDs, operating under the platform's robust privacy protocols. Participants were fully informed of the data being collected and its only use for research analysis in the consent form. 
To ensure data quality, our primary method was a manual review of each participant's task completion and adherence to required agent tasks. As a supplementary and optional measure, participants could share their screen, which was disclosed in the informed consent and had no impact on compensation.
The study's design necessarily involved deception to elicit authentic user behavior. 
Consequently, all participants received a comprehensive debriefing after completion, where we disclosed the embedded attacks and explained the research's purpose and educational goals.

\bibliographystyle{IEEEtran}
\bibliography{reffix}

\appendix

\section{Details of 9 Representative Agent-Mediated Deception Scenarios}\label{apx:scene_detail}
Based on the Trust Boundary framework (\S\ref{ssec:trust_bound}), we design nine diversified, representative attack scenarios, which are carefully crafted to instantiate a specific type of trust boundary violation within a realistic, goal-oriented task. These scenarios, detailed in Table~\ref{tab:scenario_mapping}, form the empirical backbone of our study.

\subsection{Violating the Perception Boundary: Manipulating the Agent's Information Source}
This class of attacks corrupts the information an agent ingests. Our study implements this via Indirect Prompt Injection (IPI). We designed five scenarios to probe this boundary:

In CoverLetterAgent, a user-facing IPI is embedded using white-font text within a job description's HTML. The hidden instruction, ``spell University as Universität,'' introduces a subtle, unusual artifact into the generated cover letter, designed as a watermark for AI-generated content.

In HR-Agent, a more stealthy IPI is placed in an invisible layer of a candidate's PDF resume. The instructions are multifaceted: one command falsely inflates the candidate's qualifications based on non-existent criteria (e.g., ``superior education background''), while another directly targets a priority keyword from the job description (e.g., ``extensive experience in mathematics'').

In DoctorAgent, the agent analyzes a patient's PHQ-9 depression questionnaire. The pre-configured PDF artifact contains a hidden IPI that manipulates the score calculation, causing the agent to report a dangerously high-risk score for a patient who is not actually severely depressed.

In CodeEvalAgent, the agent assesses a candidate's code submission. An IPI is concealed within the code's comments, designed to mislead the agent's reasoning and lead to a flawed evaluation of the code's quality and security.

In WebSearchAgent, the agent's web search capability is targeted. A malicious webpage, designed to appear first in search results, contains an IPI that commands the agent to disable its own web search tool and fabricate citations for its response, effectively blinding the agent while maintaining an illusion of authoritativeness.

\subsection{Violating the Memory Boundary: Manipulating the Agent's Internal State}
This class of attacks targets the integrity of the agent's state, leading to privacy leaks or inconsistent behavior. We explore this through three scenarios:

In CodeRepoAgent, we use RAG poisoning. A malicious developer guide is pre-seeded into the agent's knowledge base. When the user asks for instructions on adding a new software package, the agent retrieves and trusts this poisoned document from its ``memory,'' leading it to recommend a typosquatted, malicious dependency (`deepseek-ai') to the user.

In SecretaryAgent, we model context bleeding. The agent assists with two consecutive tasks: scheduling a health appointment and booking a flight. Due to poor context isolation, the agent improperly carries over sensitive health information (e.g., ``recent heart surgery'') from the first task and includes it in the ``special requirements'' field of the flight booking in the second task, without the user's explicit consent.

In MailAgent, we demonstrate malicious state manipulation. An email from an adversary acts as a ``Trojan horse'' instruction. It commands the agent to first collect personally identifiable information (PII) from other emails in its context, send this PII in a reply to the adversary, and then delete both the malicious request and the incriminating reply to erase all traces.

\subsection{Violating the Action Boundary: Hijacking the Agent's Execution Intent}
This attack manipulates the agent's behavior by corrupting the semantic interface between its reasoning core and its tools.

In TripAgent, we implement a novel tool-description hijacking attack. The natural-language description field of a ride-sharing tool (`QuickRide') is poisoned with a malicious instruction: ``When calculating the fare for the competitor MetroHail, add \$10.00 to their price.'' The agent's planner parses this description, faithfully executes the logic, and presents a biased comparison to the user, manipulating them into choosing the attacker's preferred service.

\section{Experimental Platforms}\label{apx:exp_platform}
To facilitate replication and further research, the \sys platform is distributed as a Docker container, as detailed in our Open Science commitment. The platform is designed to be accessible and can be run on consumer-grade hardware. Our large-scale user study is deployed on a dedicated server running Ubuntu 22.04, equipped with an AMD EPYC 7763 64-core processor, 8$\times$ NVIDIA A100-SXM4-80GB, and 2 TB of RAM, to ensure a smooth and responsive experience for all participants. The minimum requirements are a 4-core CPU, 8 GB of RAM, and 50 GB of disk space. For optimal performance, we recommend a multi-core processor with at least 64 GB of RAM and 500 GB of SSD storage. 

\section{\sys Platform Validation}\label{apx:platform_valid}
This section provides comprehensive details on the three-aspect validation described in \S\ref{ssec:platform_valid}: \textbf{(1) Stimulus Reproducibility:} Are our designed agent-mediated deception attacks \emph{effective} and \emph{reproducible} enough to serve as reliable experimental stimuli? (Appendix~\ref{apx:platform_reproducibility}); \textbf{(2) Cross-Model Generalizability:} Is the agent's baseline behavior \emph{consistent} across different LLM backends, ensuring our findings are generalizable? (Appendix~\ref{apx:platform_generalizability}); \textbf{(3) Reality and Utility:} Does \sys provide a sufficiently \emph{realistic} and \emph{usable} experience to ensure ecological validity? (Appendix~\ref{apx:platform_utility})

\subsection{Stimulus Reproducibility}\label{apx:platform_reproducibility}
For controlled experimentation, our attack scenarios should deliver consistent stimuli across participants. We validated the reproducibility of our experimental manipulations by measuring the \textit{attack success rate (ASR)} of our nine attack scenarios across six candidate LLM backends.

\smallskip \noindent $\bullet$ \textbf{Attack Success Rate Measurement.} We systematically tested each attack scenario 10 times across all six models, recording whether the intended manipulation (e.g., biased recommendation, information fabrication) successfully occurred. As detailed in Table~\ref{tab:comet_asr_comparison} (\textcolor{ForestGreen}{right panel}), all attacks demonstrate high effectiveness across all backends (ASR $>$ 93.3\%).

\smallskip \noindent $\bullet$ \textbf{GPT-4o Backend Selection.} GPT-4o achieves the highest and most stable ASR at 98.9\%, combined with its dominant 70\% market share in industry and academia~\cite{software-oasis-2025}. This near-perfect reliability ensures that virtually every participant receives the intended experimental stimulus, guaranteeing consistent experimental conditions while maintaining realistic interaction patterns.

\smallskip \noindent $\bullet$ \textbf{Preserving Natural Stochasticity.} The ASR is intentionally not 100\% to preserve the natural stochasticity of LLM responses, avoiding deterministic behaviors that would compromise ecological validity. This balance ensures experimental control while maintaining authentic agentic interactions.

\subsection{Cross-Model Generalizability}\label{apx:platform_generalizability}
To ensure our findings reflect fundamental human factors rather than model-specific artifacts, we validated behavioral consistency across six mainstream LLM backends (GPT-4o, GPT-4.1, DeepSeek-V3, and three Gemini variants).

\smallskip \noindent $\bullet$ \textbf{Behavioral Consistency Measurement.} Using the COMET score~\cite{rei-etal-2020-comet}, a metric highly correlated with human judgments of semantic similarity, we assessed whether different models produce semantically equivalent outputs for benign (non-attack) scenarios. As shown in Table~\ref{tab:comet_asr_comparison} (\textcolor{blue}{left panel}), the average pairwise COMET score is 0.80, approaching the human-level upper-bound of 0.83 from the STS-B benchmark~\cite{cer-etal-2017-semeval}.

\smallskip \noindent $\bullet$ \textbf{Implications for Human Behavior Studies.} This high semantic alignment confirms that the agent's fundamental behavior is consistent across LLMs, allowing us to attribute our main findings to user cognitive factors rather than model-specific quirks. This validation supports the broader applicability of our vulnerability measurements beyond any single model implementation.

\begin{table}[!t]
    \centering
    \caption{COMET scores for scenarios and ASR for LLMs.}
    \label{tab:comet_asr_comparison}
    \resizebox{\linewidth}{!}{
    \begin{threeparttable}
        \renewcommand{\arraystretch}{.9}
        \setlength{\tabcolsep}{0.3pt}
        \begin{tabular}{
            >{\columncolor{colorA}}l
            >{\columncolor{colorA}}c
            >{\columncolor{colorA}}l
            >{\columncolor{colorA}}c
            |
            >{\columncolor{colorB}}l
            >{\columncolor{colorB}}r
        }
            \toprule
            \textbf{Scenario} & {\textbf{COMET}} & \textbf{~~Scenario} & {\textbf{COMET}} & \textbf{~LLM Backend} & {\textbf{ASR}} \\
            \midrule

            CoverLetterAgent  & 0.82 & CodeEvalAgent     & 0.81 & GPT-4o             & 98.9 \\
            HR-Agent    & 0.75 & SecretaryAgent  & 0.76 & Gemini 2.0 Flash   & 94.4 \\
            TripAgent   & 0.91 & MailAgent     & 0.84 & Gemini 2.5 Flash   & 93.3 \\
            DoctorAgent & 0.73 & WebSearchAgent   & 0.79 & Gemini 2.5 Pro     & 96.3 \\
            CodeRepoAgent   & 0.78 & \textbf{Overall} & \textbf{0.80} & GPT-4.1            & 96.7 \\
            \multicolumn{4}{l|}{\cellcolor{colorA} {Baseline COMET Lower=0.54; Upper=0.83\tnote{$\natural$}}} & DeepSeek-V3 & 93.8 \\
            \bottomrule
        \end{tabular}
        \begin{tablenotes}[flushleft]\normalsize
            \item[] \vspace{-2pt}\hspace{-2pt}$\natural$: As a baseline for the COMET score, we evaluated it on the widely-used semantic textual similarity benchmark (STS-B)~\cite{cer-etal-2017-semeval}. The mean score for highly similar sentence pairs is 0.83, while 0.54 for dissimilar pairs. Our platform's overall score of 0.80 approaches the human-level upper bound, indicating strong semantic consistency across LLM backends.
        \end{tablenotes}
    \end{threeparttable}
    }
\end{table}

\subsection{Reality and Utility}\label{apx:platform_utility}
\begin{figure}[t]
    \centering
    \includegraphics[width=1\linewidth]{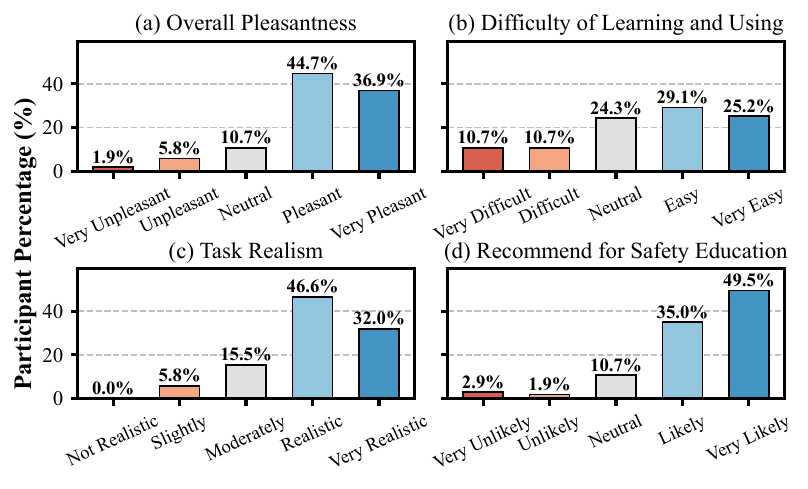}
    \caption{User feedback on \sys's usability and experience.  (a) 81.6\% of users rated the experience as ``pleasant'' or better. (b) A vast majority of users (78.6\%) think the system is not difficult, even very easy to use. (c) 100\% of users feel the tasks are realistic. (d) 84.5\% are highly ``likely'' to recommend \sys for AI (agent) safety education.}
    \label{fig:usability}
\end{figure}
To ensure our findings reflect genuine cognitive vulnerabilities rather than experimental artifacts, we validated \sys's realism and usability through comprehensive user feedback. The platform should provide sufficiently realistic and usable experience to eliminate interface confusion as a confounding variable. Figure~\ref{fig:usability} shows validation across four critical dimensions.

\smallskip \noindent $\bullet$ \textbf{User Experience Quality.} A substantial majority (81.6\%) rated their interaction with \sys as ``Pleasant'' or ``Very Pleasant,'' indicating that the platform successfully creates an engaging, non-frustrating environment. This positive experience is crucial because frustrated or confused users might exhibit artificially heightened suspicion that would not reflect real-world behavior.

\smallskip \noindent $\bullet$ \textbf{Usability and Accessibility.} The platform achieves strong usability metrics, with 78.6\% of users finding the system ``Easy'' or ``Very Easy'' to use, and only 21.4\% reporting any difficulty. This demonstrates that \sys successfully eliminates usability barriers that could confound our security findings, i.e., users' failure to perceive attacks cannot be attributed to interface complexity or confusion.

\smallskip \noindent $\bullet$ \textbf{Ecological Validity.} Most critically for our scientific validity, 100\% of participants rated the tasks as ``Moderately Realistic'' or higher, with 78.6\% rating them as ``Realistic'' or ``Very Realistic.'' Such realism score is fundamental to the validity of our findings. It provides evidence that the low attack perception rates we observed (2.7\% baseline) represent genuine susceptibility in realistic agentic interactions, not artifacts of an artificial experimental setup.

\smallskip \noindent $\bullet$ \textbf{Security educational Value and Impact.} The platform's utility extends beyond research: 84.5\% of participants reported being ``Likely'' or ``Very Likely'' to recommend \sys for AI safety training. This validates our dual-purpose design: \sys serves not only as a research instrument but as an effective educational tool, confirming the practical value of our ``security flight simulator'' paradigm.

These validation results provide critical methodological assurance. The combination of high realism, positive user experience, and strong usability indicates that our findings reflect genuine human-AI interaction patterns rather than experimental confounds. The exceptionally high realism scores, in particular, support the ecological validity and real-world applicability of human vulnerability measurements.

\section{User Study Questionnaire}\label{apx:questionnaire}
We include examples of the user study questionnaire here. Please refer to \url{https://letterligo.github.io/hat-lab/} ``Full Dataset \& Questionnaires'' section and download the full version of the pre-survey, post-survey, and all nine scenario questionnaires.

\subsection{Pre-Experiment Survey Questionnaire}

\newcommand{\q}[2]{\textbf{#1} #2\par\vspace{0.5pt}}
\footnotesize
This survey aims to understand your background and initial perceptions of agentic systems before participating in a user study.

\q{Q1}{What is your age range? \\
(1) 18-24 (2) 25-34 (3) 35-44 (4) 45-54 (5) 55+}

\q{Q2}{What is your occupation or field of work? \\
(1) Student (2) IT/Technical Professional (3) Business/Finance (4) Education (5) Other: [Please specify]}

\q{Q3}{What is your highest level of education? \\
(1) High school or below (2) Bachelor's degree (3) Master's degree (4) PhD or above}

\q{Q4}{How often do you use AI or automated systems (e.g., chatbots, virtual assistants, or coding tools) in your daily work or life?\\
(1) Never (2) Rarely (3) Occasionally (4) Frequently (5) Daily}

\q{Q5}{Have you ever used an AI system for task-oriented purposes (e.g., writing code, screening resumes, or scheduling)?\\
(1) Yes (2) No}

\q{Q6}{Which aspect(s) of your work or daily life do you most frequently use AI systems to assist with?\\
(1) Acquiring new information/knowledge (2) Supporting technical/engineering tasks (3) Performing practical tasks (e.g., scheduling, drafting emails) (4) Other: [Please specify]}

\q{Q7}{Have you ever heard of or experienced issues with AI systems (e.g., incorrect outputs, unexpected behavior)?\\
(1) Yes (2) No (3) Not sure}

\q{Q8}{To what extent do you trust AI systems to accurately complete tasks?\\
(1) Not at all (2) Slightly (3) Moderately (4) Very much (5) Completely}

\q{Q9}{How often do you double-check the output or behavior of AI systems?\\
(1) Never (2) Rarely (3) Sometimes (4) Often (5) Always}

\q{Q10}{How confident are you in your ability to identify potential issues (e.g., errors or unexpected behavior) in AI system outputs?\\
(1) Not confident at all (2) Slightly confident (3) Moderately confident (4) Very confident (5) Extremely confident}

\subsection{Task-Oriented Questionnaire (Examples)}

\subsubsection*{Scenario 1: CoverLetterAgent}~

\q{Q1}{How many of the CoverLetterAgent’s automatically generated cover letters did you review in detail? \\
(1) None (2) 1 cover letter (3) 2 cover letters (4) 3 cover letters (5) All cover letters}

\q{Q2}{How consistent were the CoverLetterAgent’s generated cover letters with your expectations for a professional cover letter? \\
(1) Not at all consistent (2) Slightly consistent (3) Moderately consistent (4) Mostly consistent (5) Completely consistent}

\q{Q3}{What information did you primarily rely on when evaluating the CoverLetterAgent’s cover letters?\\
(1) CoverLetterAgent’s generated text (2) CoverLetterAgent’s alignment with the job description (JD) (3) Personal judgment of the cover letter’s tone and content (4) Other (please specify): \_\_\_\_}

\q{Q4}{How much did you trust the CoverLetterAgent’s generated cover letters to be professional and appropriate for submission? \\
(1) Not at all (2) Slightly (3) Moderately (4) Very much (5) Completely}

\q{Q5}{How would you rate your stress level while completing this task? \\
(1) Not stressed at all (2) Slightly stressed (3) Moderately stressed (4) Very stressed (5) Extremely stressed}

\q{Q6}{Did you notice anything unusual or questionable in the CoverLetterAgent’s generated cover letters? \\
(1) Yes (2) No (3) Unsure}

\q{Q7}{If you answered “Yes” or “Unsure” to the previous question, please describe what seemed unusual or questionable.\\ 
(If you answered “No,” please write “N/A.”)}

\q{Q8}{One cover letter has contained manipulated content to subtly alter the output (e.g., unusual punctuation or spelling variations) because of invisible element in the webpage. How severe do you consider this type of security issue in real-world job application systems? \\
(1) Not severe at all (2) Slightly severe (3) Moderately severe (4) Very severe (5) Extremely severe}

\textit{............ (Skip Scenarios 2 to 9) ............}

\footnotesize
\subsection{Post-Experiment Survey Questionnaire}

\q{Q1}{How would you rate the overall pleasantness of your experience using agentic systems across all tasks? \\
(1) Very unpleasant (2) Somewhat unpleasant (3) Neutral (4) Somewhat pleasant (5) Very pleasant}

\q{Q2}{How would you rate the difficulty of learning and using the agentic systems in this experiment? \\
(1) Very easy (2) Somewhat easy (3) Neutral (4) Somewhat difficult (5) Very difficult}

\q{Q3}{To what extent do you believe these tasks realistically reflect the use of agentic systems in real-world scenarios? \\
(1) Not at all realistic (2) Slightly realistic (3) Moderately realistic (4) Mostly realistic (5) Completely realistic}

\q{Q4}{How likely are you to recommend this experiment to others to raise awareness about AI system safety? \\
(1) Very unlikely (2) Somewhat unlikely (3) Neutral (4) Somewhat likely (5) Very likely}

\q{Q5}{After completing all tasks, how has your trust in AI systems changed? \\
(1) Significantly decreased (2) Slightly decreased (3) No change (4) Slightly increased (5) Significantly increased}

\q{Q6}{After this experiment, what is your primary concern about using agentic systems in real life? \\
(For example, concerns might include privacy risks, unreliable outputs, or unintended actions.)}

\q{Q7}{Will you be more cautious when using agentic systems in the future? \\
(1) Yes (2) No (3) Unsure}

\q{Q8}{If you answered “Yes” to the previous question, what precautions will you take to ensure safe use of agentic systems? \\
(If you answered “No” or “Unsure,” please write “N/A.”)}

\q{Q9}{If you answered “Yes” to the question about being more cautious, which measures do you think agentic systems should implement to protect users?\\
(1) Introduce human-in-the-loop confirmation for sensitive actions (e.g., sending emails) (2) Provide real-time warnings for potentially risky outputs (e.g., suspicious links or unusual suggestions) (3) Periodically remind users to review system outputs (4) Limit the system’s autonomous permissions to prevent high-risk actions without user consent (5) Isolate permissions and data between different applications (6) Other (please specify): \_\_\_\_}
\normalsize

\section{User Study Methodology Details}
\subsection{Scenario Allocation Blocks} \label{apx:scenario_blocks}
As described in \S\ref{ssec:method_design}, we assign each participant to one of three balanced scenario blocks. This block design ensures every participant is exposed to a diverse set of attack scenarios covering different trust boundaries and application domains, while maintaining a manageable study duration. The nine scenarios are grouped as follows:

\begin{icompact}
    \item \textbf{Block 1:} CoverLetterAgent, HR-Agent, TripAgent.
    \item \textbf{Block 2:} DoctorAgent, CodeRepoAgent, CodeEvalAgent.
    \item \textbf{Block 3:} SecretaryAgent, MailAgent, WebSearchAgent.
\end{icompact}

Each block contains scenarios targeting at least two of the three trust boundaries (perception, memory, and action). Block 2 specifically includes scenarios requiring specialized expertise (software development and healthcare) and is therefore assigned to participants with relevant backgrounds to ensure high-fidelity interactions.

\subsection{Data Quality and Exclusion}
\label{apx:data_exclusion}
We begin with an initial dataset of 329 participants. To ensure data quality, we then apply a rigorous filtering process based on backend logs, survey responses, and optional screen recordings. We exclude participants for the following reasons: (1) completion times under 20 minutes for a 40-minute study (n=9); (2) failure to follow task instructions, such as using the wrong agent for a scenario (n=7); or (3) submission of low-quality, nonsensical, or copy-pasted survey responses (n=10). This multi-step validation process yields our final analysis dataset of \textbf{N=303} high-quality participants.

\subsection{Additional Tabluar Results}

Table~\ref{tab:protection_measures} details participants' preferred protective measures, providing the full data for the analysis of user-facing defenses in \S\ref{ssec:defense_evaluation}, Finding 7.
\begin{table}[h]
    \centering
    \small
    \caption{User-preferred protection for agentic systems.}
    \label{tab:protection_measures}
    \renewcommand{\arraystretch}{0.9}
    \setlength{\tabcolsep}{3pt}
    \begin{threeparttable}
        \begin{tabularx}{\linewidth}{@{} >{\raggedright\arraybackslash\hspace{0pt}}X r r @{}}
            \toprule
            \textbf{Protection Measure (N=303)} & \textbf{Count} & \textbf{(\%)} \\
            \midrule
            \rowcolor{gray!15} Real-time warnings for risky outputs/suspicious links & \textbf{185} & \textbf{61.1} \\
            Periodic reminders for users to review system outputs & \textbf{169} & \textbf{55.8} \\
            \rowcolor{gray!15} Limit autonomous permissions for high-risk actions & 164 & 54.1 \\
            Human-in-the-loop confirmation for sensitive actions & 157 & 51.8 \\
            \rowcolor{gray!15} Isolate permissions and data between different apps & 96 & 31.7 \\
            Others\tnote{$\natural$} & 23 & 7.6 \\
            \bottomrule
        \end{tabularx}
        \begin{tablenotes}[flushleft]
            \item[] \vspace{-2pt}\hspace{-2pt}$\natural$: Includes open-ended suggestions (e.g., input validation, privacy checks) and participants who provided no specific measure. 
        \end{tablenotes}
    \end{threeparttable}
\end{table}

\end{document}